# Typhoon intensity ReAnalysis (TyRA)

Kosuke Ito*

Disaster Prevention Research Institute, Kyoto University, Uji, Japan

Typhoon Research and Technology Center, Yokohama National University, Yokohama, Japan

Masataka Aizawa

Hokkaido University of Education Sapporo, Sapporo, Japan. Sapporo

Udai Shimada, Yasuhiro Kawabata, Munehiko Yamaguchi

Meteorological Research Institute, Tsukuba, Japan

*Corresponding author address: Kosuke Ito, Disaster Prevention Research Institute, Kyoto University. Gokasho, Uji, Kyoto 611-0011, Japan.

E-mail: ito.kosuke.2i@kyoto-u.ac.jp

ABSTRACT

Best-track data have been typically used to explore long-term changes in typhoon intensity over the western North Pacific (WNP). However, the methods used to construct best-track intensities have changed over the past several decades. Several attempts have been made to achieve homogeneous intensity analyses based on the Dvorak method, but their consistency with direct observations can still be improved. This study tries to accurately explain minimum sea-level pressure (MSLP) from aircraft observations during 1981–1986 using CI numbers and other parameters as in Aizawa et al. (2024). Using the optimized coefficients, we reanalyzed MSLP for 1987–2024. The reanalyzed MSLPs were converted into maximum wind speeds using Knaff and Zehr (2007). This Typhoon intensity ReAnalysis (TyRA) exhibited a slightly decreasing trend in the number of intense typhoons, which was not statistically significant. Compared with TyRA, the Joint Typhoon Warning Center (JTWC) best-track contained a larger number of intense typhoons from the late 1980s to the mid-2000s, whereas the Regional Specialized Meteorological Center (RSMC) Tokyo best-track contained a smaller number. In TyRA, typhoons recently weakened over the eastern part of the WNP with some statistical significance. It can be explained by the decrease of mean age of typhoons after typhoon genesis. Large differences among the RSMC Tokyo and JTWC best-track data and TyRA were found for rapid intensification. Furthermore, although previous studies reported that errors in intensity forecasts by the RSMC Tokyo increased in the mid-2000s, the changes in the quality of the RSMC Tokyo best-track data partly explain this.

## Significance statement

The goal of this study is to develop an accurate and homogeneous dataset of typhoon intensity covering the period from 1987 to 2024, using a method that is as consistent as possible with aircraft observations in the western North Pacific. Such a dataset is needed because long-term changes in typhoon intensity remain uncertain, and existing datasets have changed over time or are not fully calibrated against aircraft observations. Our dataset suggests that typhoons may have become weaker in the eastern part of the western North Pacific over the past four decades. This is likely because typhoons in that region are, on average, observed at an earlier stage, as the average genesis location has shifted northwestward recently.

# 1. Introduction

The intensity of a typhoon, namely, a tropical cyclone (TC) in the western North Pacific (WNP), is commonly represented by minimum sea-level pressure (MSLP) and maximum wind speed (Vmax). They are a key metric closely related to the severity of associated disasters. Because typhoon intensity is sensitive to sea surface temperature, global warming is expected to increase the likelihood of more intense typhoons (Knutson et al. 2020; Oouchi et al. 2006). Several studies have also reported that long-term changes in typhoon intensity may already be emerging as global warming progresses (Emanuel 2005; Mei and Xie 2016).

However, because uncertainties in typhoon intensity estimates remain substantial, careful attention must be paid to the quality of the underlying datasets. Long-term analyses of typhoon intensity over the WNP generally rely on best-track datasets produced by operational agencies such as the Joint Typhoon Warning Center (JTWC) and the Japan Meteorological Agency (JMA) serving as the Regional Specialized Meteorological Center (RSMC) Tokyo (Howell et al. 2024; Muroi 2018). Previous studies have shown, however, that long-term trends in typhoon intensity differ substantially among best-track datasets produced by different agencies (Schreck III et al. 2014; Song et al. 2010). Figure 1 shows the number of intense typhoons based on the JTWC and RSMC Tokyo best-tracks from 1978 to 2024. Here, an intense typhoon is defined as one whose lifetime minimum MSLP reached 920 hPa-equivalent or lower (see Section 3.2 for more details). The figure indicates that the estimated numbers of intense typhoons in the two best-track datasets are relatively consistent before 1987 and after the late 2000s, whereas from 1988 to the mid-2000s the number of intense typhoons in the JTWC best-track is markedly larger than that in the RSMC Tokyo best-track. Such discrepancies pose a serious problem for assessing long-term trends in typhoon intensity under climate change.

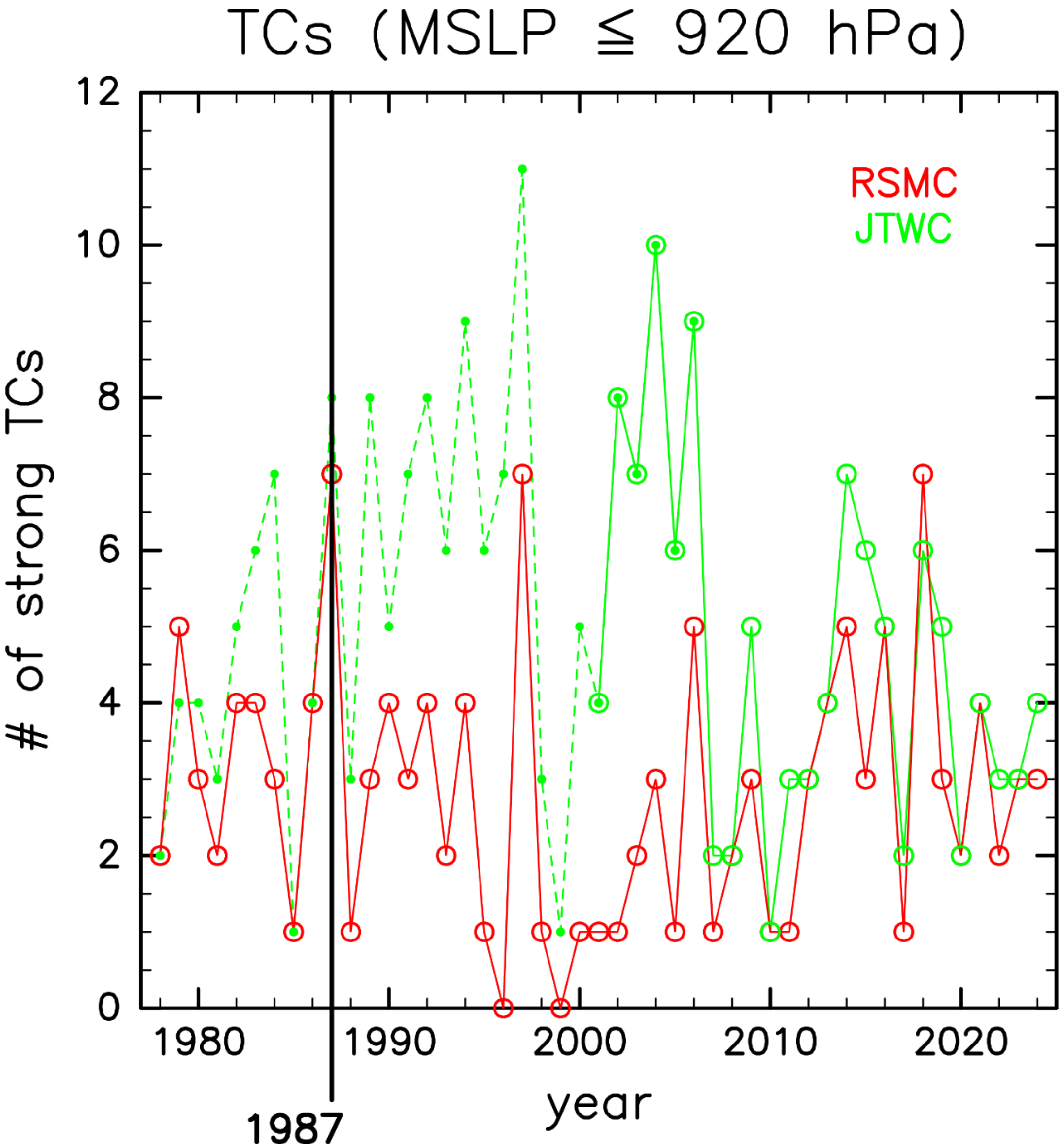

Fig. 1. Number of typhoons whose lifetime minimum MSLP reached the 920-hPa-equivalent threshold or lower. The red and green solid lines with open circles indicate the counts in the RSMC Tokyo and JTWC best-tracks, respectively. Because complete JTWC MSLP records are available only from 2002 onward, values obtained by converting Vmax to MSLP using the relationship of Atkinson and Holliday (1977) are also shown from 1978 to 2006 by the dashed line with small dots.

A major reason for this discrepancy is presumably that best-track intensities are based on the evidence and methods that each agency regarded as optimal at the time of analysis; consequently, their quality has changed over the past several decades in an agency-dependent manner. It is therefore useful to review this issue.

Because typhoons spend most of their lifetime over the ocean, accurately determining their intensity is usually difficult without direct aircraft observations. For typhoons over the WNP, intensity could be determined with relatively high accuracy from 1950s until Typhoon Cary in 1987, based on dropsonde observations of MSLP, aircraft altitudes at 700-hPa (used as the proxy of MSLP), and flight-level wind speed by frequent U.S. military reconnaissance aircraft. Both JTWC and JMA used aircraft reconnaissance information in their intensity analyses. The consistency in the number of intense typhoons between the two agencies before 1987 is likely attributable to this common observational basis. After routine aircraft reconnaissance terminated in 1987, several aircraft observation projects have been conducted over the WNP (Chan et al. 2018; D'Asaro et al. 2014; Elsberry 1990; Elsberry and Harr 2008; Hirano et al. 2022; Ito et al. 2018; Newell et al. 1996; Wu et al. 2005; Yamada et al. 2021). However, direct observations of MSLP suitable for typhoon intensity estimation have been limited to a small number of cases, such as those in TCM-90, T-PARC, ITOP, and T-PARCII (Chan et al. 2018; D'Asaro et al. 2011; Elsberry 1990; Ito et al. 2018; Weissmann et al. 2011; Yamada et al. 2021). For most typhoons, accurate intensity estimates based on direct observations are therefore still unavailable.

After the termination of routine aircraft reconnaissance in 1987, the Dvorak method came to play a central role in typhoon intensity analysis over the WNP. In this method, a tropical (T) number is derived from satellite imagery according to prescribed procedures, and a T number is converted into a current Intensity (CI) number through the adjustment depending on the typhoon stage. Finally, a CI number is converted into typhoon intensity (Dvorak 1975, 1984; Velden et al. 2006). The Dvorak method enables typhoon intensity to be estimated even in the absence of direct observations. Nevertheless, the availability of Dvorak method does not necessarily guarantee the homogeneity and accuracy in best-track intensity analyses. At least three issues can be identified: (1) determining the CI number is a difficult task, and its uncertainty affects typhoon intensity analysis; (2) biases can arise if the conversion table from CI number to intensity is inappropriate; and (3) operational agencies determine best-track intensities by considering estimates from methods other than the Dvorak method, and these procedures and weights of each information have also changed over the past several decades.

Regarding the first issue, Nakazawa and Hoshino (2009) compared T numbers and CI numbers among several organizations. They found that, during 1992–1997, JTWC CI numbers were on average approximately 0.43 higher than those of RSMC Tokyo. This may partly reflect differences in the judgment of individual forecasters (Landsea and Franklin 2013), but it is also likely attributable to differences in the detailed procedures adopted by each agency. In contrast, no large mean differences were found during the other periods examined, namely 1987–1989 and 2000–2006. This difference may explain part of the discrepancy shown in Fig. 1, but it cannot account for the intensity differences in the remaining periods and therefore cannot be the sole explanation.

Regarding the second issue, Dvorak (1984) employed the CI-number–intensity relationship of Shewchuk and Weir (1980). The table in Dvorak (1984) has been noted to produce excessively low MSLP for intense typhoons over the WNP (Kitabatake et al. 2018; Koba et al. 1991a). Shewchuk and Weir (1980) developed relationships among CI number, MSLP, and Vmax using best-track data for 51 tropical cyclones, including tropical depressions, in 1978–1979. In their formulation, the relationship of Atkinson and Holliday (1977) was used to connect MSLP and Vmax. However, that relationship did not account for the large discrepancies in sample size among different intensity categories and has been shown to assign an excessively low MSLP for a strong Vmax (Knaff and Zehr 2007; Landsea et al. 2005). In contrast, JMA related CI number to typhoon intensity using the relationship of Koba et al. (1991a), which differs from that of Dvorak (1984). With a few exceptions, Koba et al. (1991a) used only typhoons whose lifetime minimum MSLP reached 950 hPa or lower and optimized a quadratic function of CI number to match the JMA best-track. As a result, the sample sizes of direct observations were relatively balanced across different intensity categories. Knaff and Zehr (2007) and Kitabatake et al. (2018) compared the Koba et al. (1991a) conversion output with aircraft observations and showed that it is reasonable on average. Very recently, Aizawa et al. (2024) reexamined the development process of the Koba et al. (1991a) conversion relationship and demonstrated that further improvements are possible. Specifically, Aizawa et al. (2024) refined case selection, appropriately treated aircraft data, introduced JMA reanalyzed CI numbers (Nishimura et al., 2023), and incorporated additional explanatory variables such as environmental pressure and storm size into the conversion from CI number and related parameters to MSLP. As a result, they reduced the root-mean-square difference (RMSD) between aircraft-observed MSLPs and CI-number-based estimates over the WNP by more than 20% relative to Koba et al. (1991a).

Regarding the third issue, JTWC first analyzes Vmax and then estimates MSLP; since the late 2000s, the relationship of Knaff and Zehr (2007) has been referred to for relating Vmax and MSLP. Before that, the conversion was commonly done with Atkinson and Holliday (1977). Recent studies have shown that the characteristics of intensity analysis in the RSMC Tokyo best-track have also changed in the mid-2000s. Kawabata et al. (2023), using JMA reanalyzed CI numbers for 1986–2016, reported that from the late 1980s to the mid-2000s the number of intense typhoons in the RSMC Tokyo best-track was smaller than the number inferred from CI numbers, whereas after the late 2000s the two became more consistent. Shimada et al. (2020) investigated long-term changes in the number of rapid intensification (RI) cases and reported that from the late 1980s to the mid-2000s, the number of RI cases estimated from CI numbers was larger than that in the RSMC best-track, while in recent years RSMC Tokyo best-track intensity changes have become more consistent with CI-number-based estimates. These findings suggest that, from late 1980s to the mid-2000s, best-track intensity analyses by RSMC Tokyo were not necessarily based solely on the Dvorak method such as the extrapolation from ship reports using Takahashi's formula (Takahashi 1939). In addition, RSMC Tokyo traditionally estimated MSLP first and then derived Vmax (JMA 1990), whereas in recent years it has shifted to a procedure in which Vmax is estimated first and MSLP is then derived by accounting for storm size (Oyama 2022). Moreover, both JTWC and RSMC Tokyo have introduced various changes in response to new observations and methods, including the use of microwave satellite data, revised treatment of ship and surface observations, and modifications to post-analysis procedures.

Although such methodological or procedural updates are believed to contribute to improved accuracy in typhoon intensity analysis, these changes make it difficult to detect accurate long-term changes in typhoon intensity from best-track data. Given these circumstances, accurately capturing long-term trends in typhoon intensity requires reconstruction of MSLP and Vmax using methods that are as homogeneous and accurate as possible.

Recent efforts have made progress toward such typhoon intensity reanalysis. Kossin et al. (2020) developed ADT-HURSAT, a global TC intensity reanalysis dataset based on the Advanced Dvorak Technique (ADT), an automated system for estimating TC intensity from satellite imagery, and used it to analyze global TC intensity. They found that mean global TC intensity has generally strengthened, although the intensification over the WNP was relatively modest. In addition, following the emphasis placed on homogeneous typhoon intensity analysis at the Seventh International Workshop on Tropical Cyclones (WMO 2011), RSMC Tokyo conducted a Dvorak-based reanalysis of CI numbers for typhoons from 1987 to 2016 (Nishimura et al. 2023). Using these homogeneously analyzed CI numbers, Kawabata et al. (2023) showed that there was no significant long-term trend in the number of intense typhoons over the western North Pacific as a whole. Sezaki et al. (2025) also showed, based on the Dvorak method, that the number of violent typhoons exhibited no long-term trend.

These developments are valuable for accurately assessing the influence of climate change and for understanding environmental controls on typhoon intensity. Nevertheless, the CI–MSLP relationship used in ADT-HURSAT is basically that of Dvorak (1984), and thus may retain the biases discussed above. Studies based on JMA reanalyzed CI numbers, such as Sezaki et al. (2025), are expected to provide reasonably accurate estimates. However, the conversion relationship of Aizawa et al. (2024), which better explains reliable aircraft observations over the WNP, may yield more accurate intensity estimates. Also, geographical distribution and other characteristics should be investigated. Furthermore, although the JMA CI-number reanalysis is limited to 1987–2016, JMA post-analysis CI numbers allow typhoon intensity analysis to be extended over a longer period until very recently.

In this study, we formulate a relationship based on Aizawa et al. (2024) that accurately explains aircraft-observed typhoon MSLP over the WNP during 1981–1986 using CI numbers and other basic variables, and apply it to estimate MSLP for 1987–2024. We then derive Vmax speed using the relationship of Knaff and Zehr (2007). Through this procedure, we construct a freely accessible Typhoon intensity ReAnalysis (hereafter referred to as TyRA; http://hdl.handle.net/2433/301586), consisting of MSLP and Vmax. This approach enables a long-term intensity evaluation that is relatively homogeneous and accurate.

The TyRA dataset developed in this study can be used for a wide range of applications, including appropriate assessment of climate-change impacts, development of future global and regional reanalysis datasets, evaluation of forecast experiments, inputs for hindcasts by ocean and wave models, and investigation of environmental factors controlling typhoon intensity. In this paper, we describe the procedure used to construct TyRA, evaluate its accuracy, and conduct basic analyses of long-term changes in typhoon intensity and related quantities. The remainder of this paper is organized as follows. Section 2 provides an overview of the procedure used to construct the reanalysis. Section 3 presents long-term changes in typhoon intensity, their geographical distribution, rapid intensification rate, and intensity forecast errors based on TyRA. Section 4 summarizes the main conclusions.

# 2. Method

*2.1 Overview of the reanalysis procedure*

Figure 2 shows an overview of the procedure used to construct the TyRA. The basic dataset used for optimization consists of aircraft observations of intense typhoons conducted by JTWC during 1981–1986. The coefficients were determined using the functional form of TEST1 in Aizawa et al. (2024), in which MSLP is modeled by considering the square of the CI number, the CI number, the change in the CI number in the last 24 h, typhoon size, the latitude of the typhoon center, and the environmental sea-level pressure. The optimized coefficients were then used to estimate MSLP from CI numbers and related variables for 1987–2024. The estimated MSLP was converted into 1-min mean Vmax using the relationship of Knaff and Zehr (2007). Through this procedure, a 6-hourly typhoon intensity reanalysis was constructed for 1987–2024.

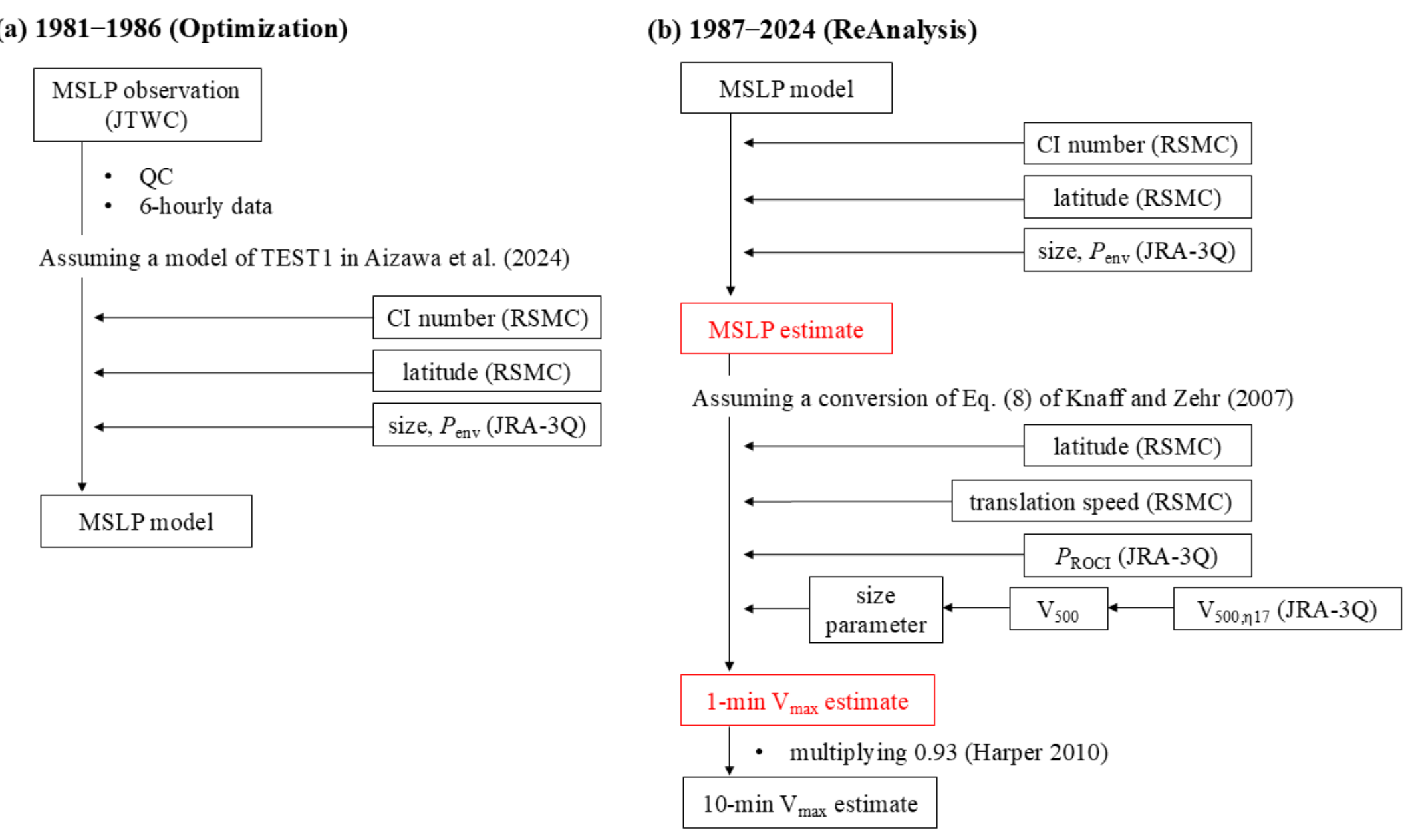

Fig. 2. Overview of the procedure used to construct the reanalysis dataset.

Many operational agencies first estimate Vmax and then convert it into MSLP. In contrast, this study first derives MSLP and then converts it into Vmax. This choice is motivated by the fact that the most reliable direct observations available for WNP typhoon intensity are aircraft-observed MSLPs. In WNP aircraft reconnaissance in 1980s, Vmax was not directly observed; at most, it was inferred from visual observations of sea-surface conditions or from flight-level winds. Vmax values recorded in JMA at that time were generally determined by emphasizing conversions from MSLP, such as the Takahashi formula (Aizawa et al. 2024; Takahashi 1952), $V_{\max} = \sqrt{1010 - MSLP}$ . At JTWC, the relationship of Atkinson and Holliday (1977), $V_{\max} = 6.7\left(1010 - MSLP\right)^{0.644}$, was commonly emphasized. Knaff and Zehr (2007) and Courtney and Knaff (2009) showed that MSLP and Vmax can be related with high accuracy by accounting for sample biases, latitude dependence, size parameters, typhoon translation speed, and environmental pressure. Their results were obtained mainly from Atlantic and Gulf of Mexico datasets, where many aircraft observations using various instruments for MSLP and Vmax are available, and are therefore considered to have a certain degree of reliability. Because reliable wind-speed observations are available for only a very limited number of cases in the WNP, TyRA adopts an approach in which optimization is based on MSLP, the most reliable observed quantity, and Vmax is then calculated by converting the optimized MSLP using Knaff and Zehr (2007). This approach is also beneficial in terms of disaster prevention and mitigation as Klotzbach et al. (2020) demonstrated that MSLP is a better matric than Vmax.

*2.2 Aircraft observation data*

This subsection describes the aircraft-observed MSLP data from 1981 to 1986 used as the target for optimization over WNP. Aircraft observation data related to MSLP for 1981–1985 were taken from JTWC annual reports. The data for 1986 were not included in the texts of the reports but were recorded on floppy disks distributed by JTWC at that time. The 1986 data were provided by JTWC for this study. The JMA Geophysical Review also contains MSLP values based on dropsondes and those estimated from flight-level altitude. In principle, MSLP values from the JTWC annual reports were used. However, when the JTWC annual reports contained obvious typographical errors or unclear entries, the JMA Geographical Review that also recorded the aircraft reconnaissance was also referenced. When values in both sources could not be judged as clearly erroneous but were mutually inconsistent, the data were not used.

Following Aizawa et al. (2024), this study used MSLP data for 51 typhoons during 1981–1986, covering stages from their genesis to weakening. These 51 typhoons consist of 50 typhoons whose lifetime minimum MSLP was 950 hPa or lower, plus Typhoon Judy in 1982, whose lifetime minimum MSLP was 955 hPa. This case selection is nearly identical to that of Koba et al. (1991a), except that Typhoon Doyle in 1984. The lifetime minimum MSLP of Typhoon Doyle (1984) was 940 hPa, but it was omitted in Koba et al. (1991a) but included here. The reason why the 950-hPa criterion was not applied strictly in some cases in Koba et al. (1991a) is unclear. Despite these exceptions, the selected cases basically cover the genesis-to-weakening stages of typhoons with low lifetime minimum MSLP, resulting in a relatively balanced sample size across different intensity categories. This approach effectively avoids the sample size inhomogeneity by intensity category seen in Atkinson and Holliday (1977) (Knaff and Zehr 2007).

Aircraft-based MSLP estimates were obtained using dropsonde observations, values converted from 700-hPa flight-level altitude, or both. As shown in Fig. 3b of Aizawa et al. (2024), when both types of data are available, MSLP estimated by applying the Jordan (1958) formula to 700-hPa flight-level altitude agrees very well with dropsonde-based MSLP (RMSD of 3.5 hPa). Therefore, in this study, when only flight-level altitude was available, MSLP converted from flight-level altitude using Jordan (1958) was regarded as the observed value. However, data with 700-hPa altitude exceeding 3400 m were excluded because such values were clearly unrealistic and likely to be typographical error. When both dropsonde observations and 700-hPa flight-level data were available, their average was adopted as MSLP. In rare cases where both dropsonde-based and flight-level-based MSLP values were available but differed unrealistically by 10 hPa or more, both values were excluded. After the quality control described above, the RMSD between dropsonde-based and altitude-based MSLPs was reduced to 2.0 hPa for cases when both data were available.

As in best-track datasets, this study aims to construct a 6-hourly dataset at 0000, 0600, 1200, and 1800 UTC. Therefore, the value from the observation within 3 h of an analysis time was used. When multiple aircraft observations were available within 3 h, we employed the one obtained at closest to the analysis time. If aircraft observations indicated an abrupt MSLP change of 20 hPa or more within 6 h, both values were excluded as unrealistic. After applying the procedures described above, 873 aircraft-observed MSLP data from 1981–1986 were employed.

*2.3 CI numbers*

CI number is a key variable for constructing TyRA. For the optimization period of 1981–1986, the optimization of TyRA uses the CI-number reanalysis of Tokuno et al. (2009), covering all 51 typhoons described in the previous subsection. Meanwhile, Tokuno et al. (2009) did not conduct CI-number reanalysis for the other typhoons during 1981–1986. For 1987–2016, the reanalyzed CI numbers of Nishimura et al. (2023) were used, and for 2017–2024, JMA post-analysis CI numbers were used.

The CI-number reanalysis of Tokuno et al. (2009) includes two versions: a constrained version, in which abrupt changes in the T number are suppressed, and an unconstrained version, in which no such constraint is applied. The constrained version refers to the operational constraints used by JMA, under which changes in the T number must remain within ±1.0 over 6 h, ±1.5 over 12 h, ±2.0 over 18 h, and ±2.5 over 24 h (Koide and Nishimura 2017). Because the analyses of Nishimura et al. (2023) and the JMA post-analysis CI numbers were conducted using the constrained procedure, the constrained version of Tokuno et al. (2009) was also adopted here, unlike Aizawa et al. (2024) that employed the unconstrained version.

For rapid intensification, the constraint smooths some time periods in approximately 10% of the 51 typhoons and thus it serves to underestimate the number of RI events. During rapid weakening near landfall, the CI number may fail to keep pace with the actual weakening. JMA adjusts T numbers and CI numbers after typhoons pass over land areas such as the Philippines (Koba et al. 1991b), and a certain degree of accuracy is therefore expected in such cases. Nevertheless, because the current version of TyRA is constructed using an optimization based on MSLP obtained from aircraft observations over the ocean, future improvement may be possible by applying corrections to typhoon intensity over land.

In summary, three types of CI numbers were used to construct TyRA: the reanalysis of Tokuno et al. (2009), the reanalysis of Nishimura et al. (2023), and the JMA Dvorak-based post-analysis values. Although CI numbers can vary depending on the analyst who performs the analysis, these three sets of CI numbers were analyzed by experienced forecasters based on JMA modern procedures. They are not strictly identical, but they are considered to provide relatively homogeneous analyses.

The first Himawari satellite, launched in 1978, had resolutions of 1.25 km in the visible channel and 5 km in the infrared channel, which are coarser than those of the current Himawari-8/9 satellites, namely 0.5 km in the visible channel and 2.0 km in the infrared channel. However, except for some cases with extremely small eyes, these resolutions are sufficient for identifying the basic structure of typhoons. Therefore, the influence of differences in satellite image quality on TyRA is considered to be minor.

*2.4 Conversion formula from CI numbers to MSLP*

In this study, aircraft-observed MSLP during 1981–1986 was modeled using the TEST1-type equation of Aizawa et al. (2024).

$$\mathrm{MSLP} = a_1\mathrm{CI}^2 + a_2\mathrm{CI} + a_3\delta\mathrm{CI}_{24} + a_4\Phi + a_5R + a_6 + P_{\mathrm{env}} \quad (1)$$

Here, $\delta\mathrm{CI}_{24}$ is the change in the CI number over the past 24 hours, $\Phi$ is the latitude of the typhoon center, $R$ is the typhoon size in km, and $P_{\mathrm{env}}$ is the environmental sea level pressure. MSLP is mainly explained by the first two terms in the right-hand side, while other terms serve to further enhance the quality of the dataset.

As described above, the aircraft observations used to determine the coefficients were obtained for 51 typhoons mentioned in section 2.2. The sample size therefore does not vary much among CI number categories. Nevertheless, the sample size is somewhat limited for very intense typhoons with CI numbers of 7.0 and 7.5 (Fig. 3). Therefore, in this study, the coefficients $a_1, a_2, \ldots, a_6$ were determined using weighted least squares for the following function, where the weight $w_i$ is defined as the inverse of the frequency of each CI number category.

$$J = \sum_{i=1}^{873} w_i \left( P_{\min,i} - a_1 \mathrm{CI}_i^2 - a_2 \mathrm{CI}_i - a_3 \delta \mathrm{CI}_{24,i} - a_4 \Phi_i - a_5 R_i - a_6 - P_{\mathrm{env}} \right)^2 \quad (2)$$

To calculate $\delta\mathrm{CI}_{24}$ in Eq. (1), the CI number 24 h before the analysis time was subtracted from the CI number at the analysis time. When no value was available, $\delta\mathrm{CI}_{24}$ was set to 0. For this calculation, early-stage Dvorak values for tropical depressions below 34 kt were also used. This was done to avoid a large discontinuity in typhoon intensity analysis between 18 h and 24 h after formation. The latitude $\Phi$ was taken from the typhoon center position in the RSMC Tokyo best-track.

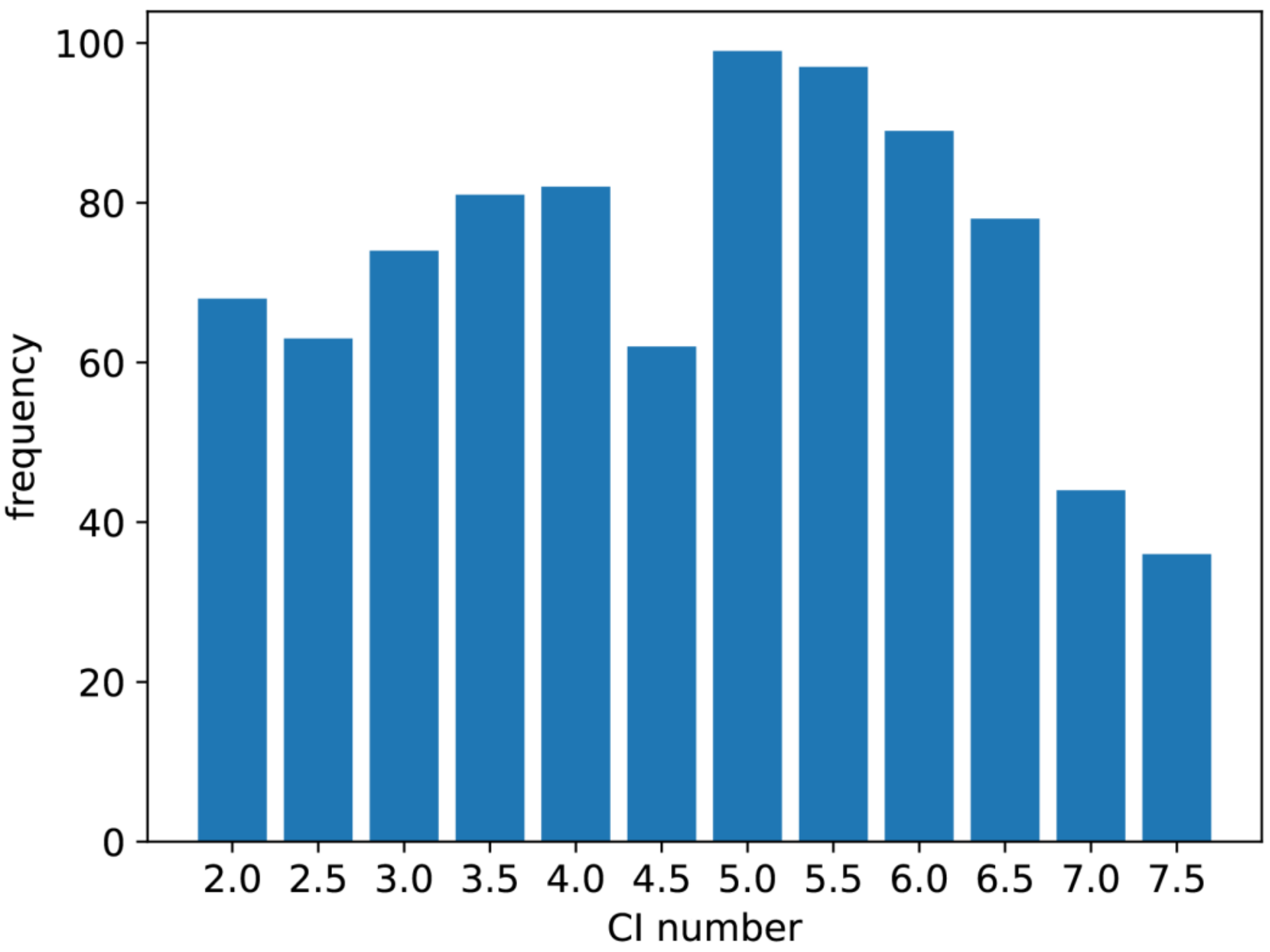

Fig. 3. Frequency distribution of CI numbers for the 873 cases.

The typhoon size $R$ was calculated using model-level data from the JMA global atmospheric reanalysis dataset JRA-3Q (Kosaka et al. 2024). Horizontal grid spacings are approximately 40 km. First, within 2.5° of the center position in the RSMC Tokyo best-track, the grid point with the MSLP was defined as the typhoon center. Then, $R$ was defined as the maximum radius, in km, at which the azimuthally averaged wind speed at the 17th vertical level (corresponding approximately to 850 hPa when the surface pressure is 1000 hPa) exceeded 15 m $s^{-1}$. If no radius with an azimuthally averaged wind speed exceeding 15 m $s^{-1}$ existed, or if $R < 100$ km, $R$ was set to 100 km. The environmental pressure $P_{\mathrm{env}}$ was calculated as the mean sea-level pressure at grid points located between $R$+100 km and $R$+300 km from the center.

The optimized equation obtained by fitting Eqs. (1) and (2) to aircraft-observed MSLP (in hPa) during 1981–1986 is as follows.

$$P_c = -2.318\mathrm{CI}^2 + 7.044\mathrm{CI} + 0.8835\delta\mathrm{CI}_{24} - 0.3809\Phi - 2.560\times10^{-2}R - 6.046 + P_{\mathrm{env}} \quad (3)$$

The coefficients indicate that an MSLP is largely dependent on a CI number, while it is adjusted by the temporal change of the CI number, typhoon size, latitude, and environmental pressure. The RMSD between the aircraft observations and the values from this equation was 9.79 hPa during the optimization period from 1981 to 1986.

*2.5. Conversion from MSLP to Vmax*

As described in the Introduction, this study obtains Vmax by converting the MSLP estimated through the procedure in Section 2.1 using the relationship of Knaff and Zehr (2007). The relationship is expressed as follows.

$$V_{\max1} = 18.633 - 14.960S - 0.755\Phi - 0.518\left(P_c - P_{ROCI}\right) + 9.738\sqrt{\left|P_c - P_{ROCI}\right|} + 1.5c^{0.63} \quad (4)$$

Here, $V_{\max1}$ is the 1-min averaged Vmax in kt, $S$ is the size parameter, $P_{\mathrm{ROCI}}$ is the pressure in hPa obtained by adding 2 hPa to the sea-level pressure at the outermost closed surface isobar, and $c$ is the translation speed in kt. The size parameter S is defined as follows.

$$S = V500 / V500_C \quad (5)$$

Here, $V_{500}$ is the surface wind, in kt, averaged within 400–600 km from the typhoon center. $V_{500C}$, which corresponds to the climatological value of $V_{500}$, is expressed as follows.

$$V500_C = V_{\max 1}\left[\frac{66.785 - 0.09102V_{\max 1} + 1.0619\left(\Phi - 25\right)}{500}\right]^{(0.1147 + 0.0055V_{\max 1} - 0.001(\Phi - 25))} \quad (6)$$

For the calculation of $V_{500}$, surface-wind-equivalent values were derived from the wind speed at the 17th vertical level of JRA-3Q (corresponding approximately to 850 hPa when the surface pressure is 1000 hPa) using the following equation.

$$V500 = 0.584 \times V500_{\eta 17} + 1.79 \quad (7)$$

This relationship was obtained by least-squares fitting between the wind speed at the 17th vertical level and the surface wind speed for 51 typhoons during 1981–1986, using cases in which $V_{500}$ could be calculated entirely over the ocean. A surface-wind-equivalent value converted from the wind speed at the 17th vertical level, approximately 850 hPa, was used rather than the simple surface wind in order to prevent underestimation of the over-ocean $V_{500}$ when a typhoon was approaching land. However, when the typhoon center is located sufficiently far inland, this approach may overestimate the Vmax actually observed. Future improvement may be possible by correcting TyRA wind speeds over land using in situ observations. The radius of the outermost closed isobar, ROCI, was identified using the algorithm described in Section 5.2 of Aizawa et al. (2024), and the outermost closed contour was adopted. The search range for ROCI was limited to within 3000 km. When another typhoon or low-pressure system was present nearby, the outermost contour was searched for in the outer region enclosing both low-pressure systems. The latitude was taken from the RSMC Tokyo best-track. The typhoon translation speed was calculated from the difference between the current and 6-h-prior positions in the JMA best-track.

Because $V_{\max}$ depends on $S$ in Eq. (5), while $S$ depends on $V_{\max}$ in Eq. (6), the solution was obtained iteratively. The $V_{\max}$ calculated from Eq. (5) with $S$=1.0 was used as the first guess. Following Courtney and Knaff (2009), the minimum value of S was set to 0.4. In very rare cases (5 records in the 38-year analysis), the estimated 1-min mean Vmax fell below the 34-kt threshold for typhoons. In TyRA, these values were set to 34 kt. $V_{\max}$ in TyRA corresponds to a 1-min mean value. When it was necessary in this paper to exchange between 10-min averaged $V_{max}$ and a 1-min averaged $V_{max}$, the following conversion was applied following Harper et al. (2010).

$$V_{max1} = 0.93 \times V_{max10} \tag{8}$$

The authors consider that obtaining reliable statistical records of both 1-min and 10-min mean wind speeds over the ocean in the WNP is extremely difficult, and that the uncertainty associated with this conversion is therefore very large.

*2.5. Reanalyzed cases*

TyRA is based on the regression equation in Eq. (3), and the analyzed cases correspond to the 6-hourly typhoon intensity analyses of 958 typhoons in the RSMC Tokyo best-track for 1987–2024. During the reanalysis period from 1987 to 2024, MSLP values for 19,925 records were obtained by inputting CI number, latitude, $R$, and $P_{env}$ into Eq. (3). The analyzed cases mostly correspond to typhoons in the RSMC Tokyo best-track with 10-min averaged $V_{max}$ of 34 kt or higher. For 0.5% of the cases, such as those immediately before completion of extratropical transition, no CI numbers from the JMA reanalysis or post-analysis are available. Therefore, although these typhoon intensity records are included in the RSMC Tokyo best-track, they are not currently included in TyRA. We also included the intensity analysis for 51 TCs during the optimization period using Eq. (3) from 1981 to 1986, meaning that the analysis was made only for the lifetime minimum MSLP was 950 hPa or lower except typhoon Judy (1982).

TyRA aims to provide a long-term homogeneous dataset rooted in consistency with aircraft observations obtained over the WNP. It is suitable for the long-term analysis. Nevertheless, satellite observations remain the primary basis for estimation, and TyRA does not necessarily provide the optimal estimate for every case. For example, even when in situ observations are available near the typhoon center and a highly reliable intensity estimate could be obtained from them, TyRA still estimates intensity using satellite-derived parameters. In this sense, TyRA should not be regarded as the optimal estimate for all individual cases. Improving the accuracy of typhoon intensity estimates using information sources over land and near coastlines remains an important task for future work.

## 3. Results

### *3.1 Optimization and verification*

We first examine the consistency of the fitted MSLP estimates during the optimization period of 1981–1986. The RMSD between the aircraft observations and the modeled values during the fitting period was 9.79 hPa. For comparison, the conversion relationship from CI number to MSLP, based on Koba et al. (1991a), was also tested:

$$P_c = -1.53\mathrm{CI}^2 - 3.03\mathrm{CI} + 1010.01 \quad (9)$$

When this equation was applied, the RMSD between the modeled MSLP and aircraft-observed MSLP was 12.58 hPa. Thus, the equation adopted in TyRA reduced the residuals by approximately 20%. On the other hand, the RMSD of 9.79 hPa is slightly larger than the value of 9.34 hPa obtained by Aizawa et al. (2024) using unconstrained T number data. Aizawa et al. (2024) reported that the use of constrained T number data increased the RMSD by approximately 0.4 hPa, and this difference is likely the main reason for the slightly degraded accuracy.

Figure 4 compares aircraft-observed MSLP with the values estimated using the proposed regression equations. Fig. 4a shows the results obtained using Eq. (9) that depends only on the CI number, whereas Fig. 4a shows the results obtained using the TEST1-type regression equation in TyRA. Comparison of three panels indicates that the TEST1-type regression equation of Aizawa et al. (2024) is more consistent with the aircraft observations, compared to the conversion of Koba et al. (1991a) and ADT-HURSAT of Kossin et al. (2020)  in which a latitude-dependent correction to the CI-number output was applied (NOAA/NCEI 2026). It should be noted that TyRA tends to underestimate the intensity of some very intense typhoons with MSLP below 920 hPa, particularly with MSLP below 900 hPa. It is therefore necessary to recognize this limitation in TyRA when estimating the intensity of extremely intense typhoons.

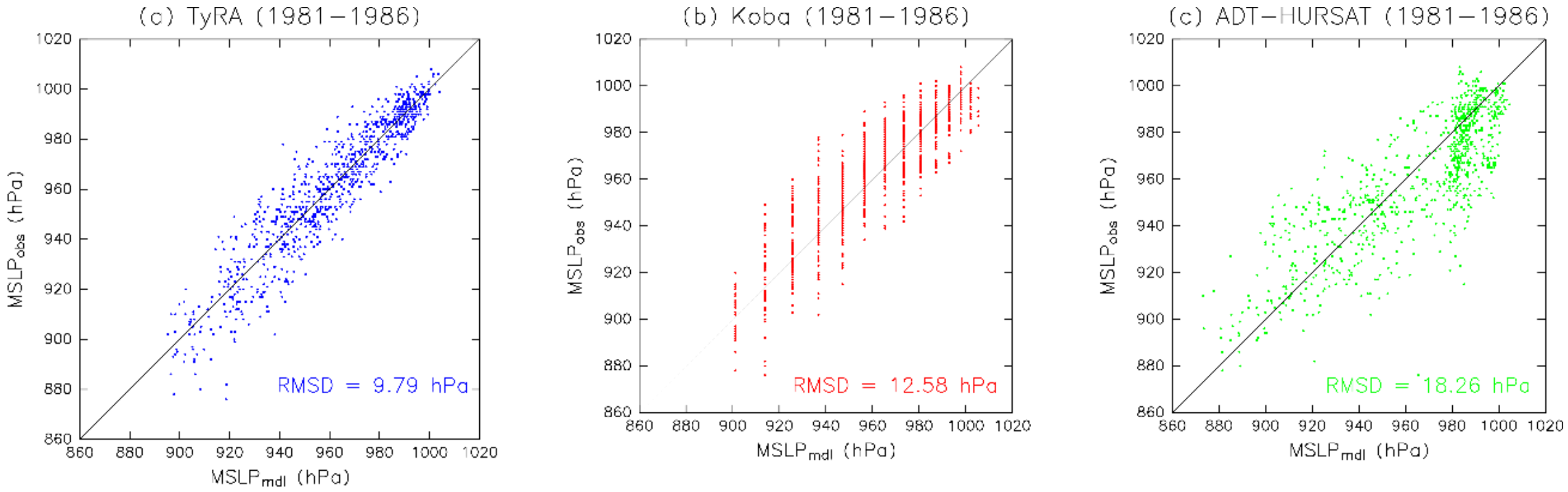

Fig. 4. Comparison between aircraft-observed and model-estimated MSLPs during the optimization period from 1981 to 1986: (a) TyRA, (b) Koba, and (c) ADT-HURSAT.

Next, Fig. 5 shows mean bias and RMSD of the deviations in estimated MSLPs from aircraft observations, defined as $\Delta\text{MSLP} = \text{MSLP}_{mdl} - \text{MSLP}_{obs}$, for each category of modeled MSLP. In addition to TyRA and Koba et al. (1991a), the figure also shows MSLP values obtained by applying the CI numbers of Tokuno et al. (2009) to the Dvorak (1984) relationship, as well as MSLP values from ADT-HURSAT of Kossin et al. (2020). The results indicate that, as described in the Introduction, Dvorak (1984) has a positive intensity bias for very intense typhoons. In particular, for cases in which the modeled MSLP is between 890 and 930 hPa, Dvorak (1984) estimates MSLP values that are, on average, approximately 20 hPa lower than the aircraft observations. ADT-HURSAT also tends to overestimate the intensity of very intense typhoons and underestimate the intensity of weak-to-moderate typhoons. As noted above, Kitabatake et al. (2018) reported that the Dvorak (1984) table tends to overestimate the intensity of intense typhoons over the WNP relative to aircraft observations. Although ADT-HURSAT applies a latitude correction, a bias in MSLP still remains. When the reanalyzed CI numbers are converted using the equation of Koba et al. (1991a), the bias is substantially reduced, consistent with Knaff and Zehr (2007) and Kitabatake et al. (2018). Nevertheless, Koba et al. (1991a) still has an average bias of approximately 4 hPa toward overly strong intensity for typhoons with MSLP of 910–970 hPa, and an average bias of approximately 5 hPa toward overly weak intensity for typhoons with MSLP of 990–1010 hPa. In TyRA, these biases are largely removed, and the mean bias is small. Although TyRA has a weak-intensity bias of approximately 2 hPa on average in the 890–910 hPa category, it achieves high accuracy on average in all other categories, and its RMSD is substantially smaller than that of the other products.

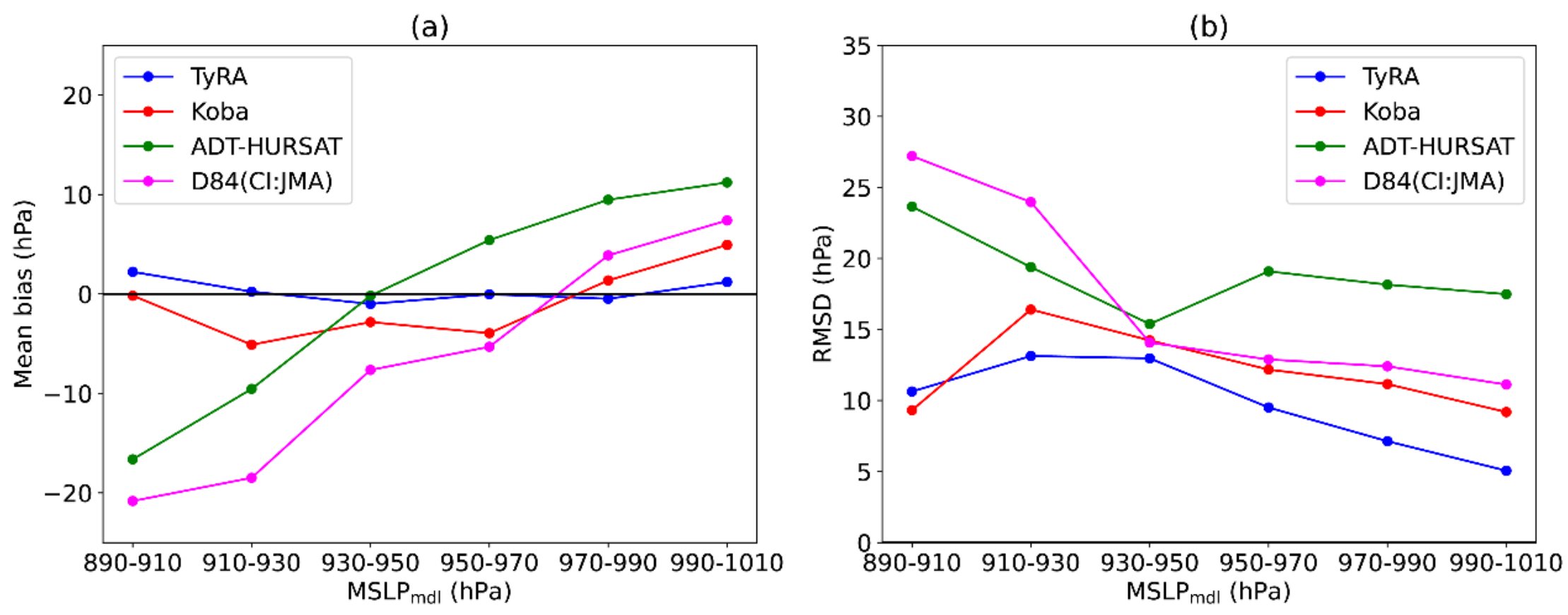

Fig. 5. Statistics of the MSLP estimates for each model-estimated MSLP category: (a) mean of $MSLP_{mdl}$ - $MSLP_{obs}$, and (b) RMSD.

Intensity bias depends both on the accuracy of the CI-number assignment and on the relationship between the CI number and MSLP. To examine this issue in more detail, Fig. 6 shows the mean MSLP for each CI number. The differences between the CI numbers assigned by ADT and RSMC Tokyo are particularly pronounced for CI numbers of 2.0 or 2.5. For example, among the 63 cases in which the ADT CI number was 2.0 or 2.5, 25 cases had aircraft-observed MSLP below 970 hPa. This indicates that relatively strong typhoons were assigned low CI numbers by ADT. In contrast, among the 131 cases in which the RSMC Tokyo CI number was 2.0 or 2.5, only 4 cases had aircraft-observed MSLP below 970 hPa. For CI numbers of 3.0 or higher, differences between the mean aircraft-observed MSLP values corresponding to each CI-number category in ADT and Tokuno et al. (2009) are smaller than those for CI numbers of 2.0 or 2.5. Even so, ADT tends to assign slightly lower MSLP values for the same CI number. Compared with aircraft observations, ADT-HURSAT tends to evaluate weak typhoons with small CI numbers as too weak, and intense typhoons with CI numbers of 6.0 or higher as too strong by assigning excessively low MSLP values. In particular, for typhoons with CI numbers of 7.0 and 7.5, the bias was approximately 15 hPa. Thus, for 1981–1986, the MSLP values output by ADT-HURSAT contain intensity-dependent biases relative to aircraft observations. The ADT-HURSAT results shown here are after latitude correction; without the latitude correction, the bias for intense typhoons becomes approximately 3 hPa larger. This result is consistent with Fig. 5 and suggests that the main cause of the bias of ADT-HURSAT relative to aircraft observations in the 1980s is likely to be the use of the Dvorak (1984) conversion relationship, which assigns excessively low MSLP values to CI numbers of 6.5 or higher. The bias in the Koba relationship is generally small, but TyRA is even more consistent with the aircraft observations evaluated using RSMC CI numbers, demonstrating its high accuracy with respect to the aircraft observations from 1981 to 1986.

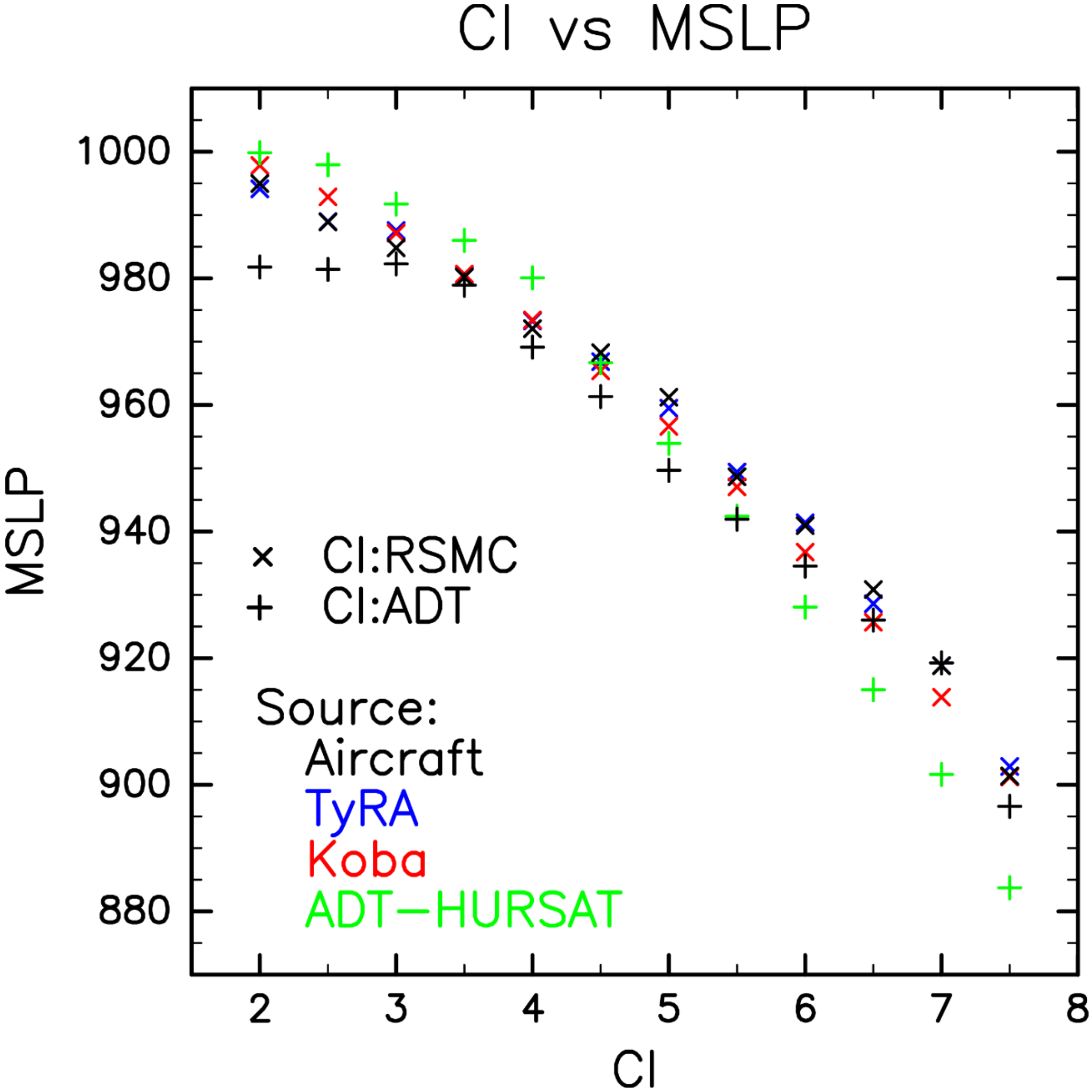

Fig. 6. Mean MSLP for each CI number in the optimization cases. Results based on RSMC CI numbers are shown by circles, whereas those based on ADT CI numbers are shown by plus signs. Black, blue, red, and green indicate aircraft observations, TyRA, Koba, and ADT-HURSAT, respectively.

In addition to the aircraft observations from 1981 to 1986 used for the optimization period, a small number of aircraft-observed MSLP data are also available for recent periods included in the reanalysis. These observations were obtained during T-PARC in 2008,

ITOP in 2010, and T-PARCII from 2017 to 2025; observations from 2025 were excluded because they fall outside the reanalysis period. Comparisons were made for 41 six-hourly best-track records using aircraft observations within ±3 h of each analysis time. The RMSD was 9.30 hPa for the Koba table and 10.27 hPa for TyRA (Fig. 7). Because the sample size is limited, particularly for very intense typhoons, it is difficult to determine whether Koba et al. (1991a) or TyRA performs better. For example, if only the two cases for Typhoon Megi on 17 October 2010 are excluded, the RMSD values become 8.89 hPa and 8.93 hPa, respectively, making the two methods nearly equivalent. Furthermore, when the JMA reanalyzed CI numbers were replaced with automatically derived ADT CI numbers, the RMSD values were 10.15 hPa for the Koba table and 9.68 hPa for TyRA, reversing the relative performance of the two methods (Fig. 8). More direct observational samples, especially for typhoons with MSLP below 900 hPa, and further improvements in CI-number estimation are needed to evaluate more accurately which of the two methods is superior. For reference, the RMSD between ADT-HURSAT MSLP estimates and aircraft observations during this period was 13.31 hPa. Although the bias in ADT-HURSAT estimates for typhoons with MSLP of 920 hPa or lower appears to be small, further verification is needed. In any case, recent direct observations support that the MSLP errors in the Koba table and TyRA are approximately 10 hPa.

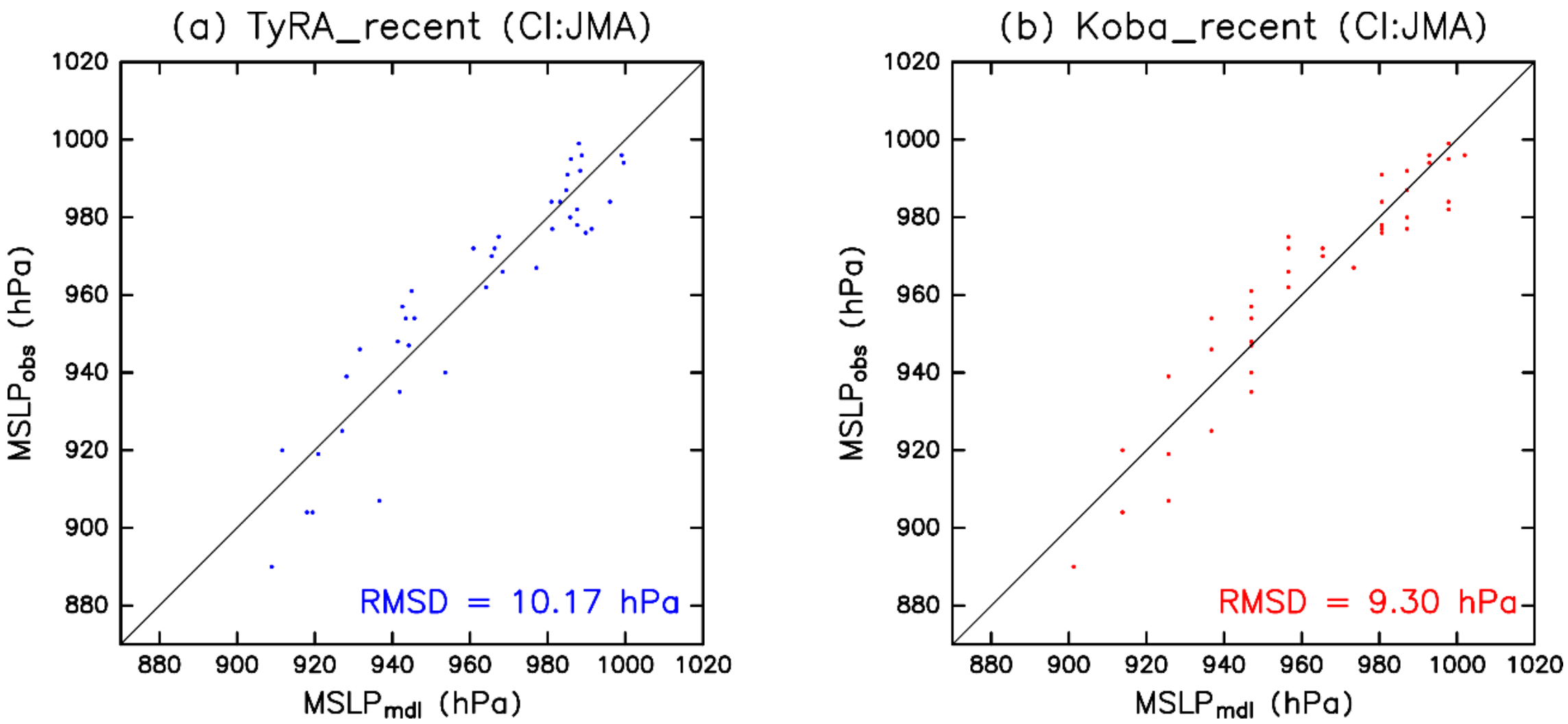

Fig. 7. Comparison of model-estimated MSLP with values obtained from recent aircraft observations.

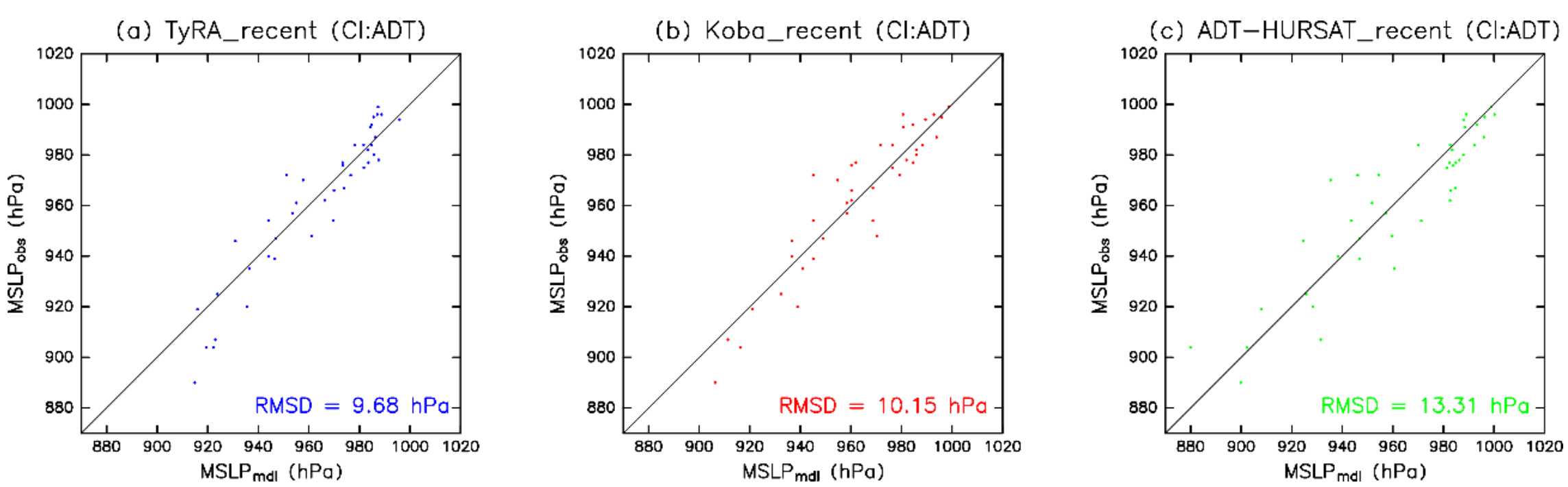

Fig. 8. Same as Fig. 7, but with the JMA-estimated CI numbers replaced by ADT-estimated CI numbers. Results from ADT-HURSAT are also included in (c).

### *3.2 Typhoon intensity*

Figure 9 shows the number of intense typhoons from 1981 to 2024. Here, intense typhoons are counted using two criteria: MSLP and Vmax. For MSLP, the threshold is a lifetime minimum MSLP of 920 hPa, corresponding to a CI number of 6.0, which was used as the criterion for intense typhoons by Kawabata et al. (2023). For Vmax, the threshold is a lifetime maximum 1-min mean Vmax of 130 kt, corresponding to the super-typhoon criteria. Note that best-track datasets usually record strong typhoon intensities rounded to the nearest 5 hPa or 5 kt. Therefore, in TyRA, typhoons with lifetime minimum MSLP of 922.5 hPa or lower and lifetime maximum Vmax of 127.5 kt or higher are counted as typhoons that would be rounded to 920 hPa and 130 kt, respectively. These are referred to here as the 920-hPa-equivalent and 130-kt-equivalent thresholds. For JTWC MSLP, complete records are available only from 2002 onward. Therefore, JTWC MSLP was evaluated using values converted from Vmax with the relationship of Atkinson and Holliday (1977). This relationship was used by JTWC to relate Vmax and MSLP until the mid-2000s. For 2002–2006, the number of intense typhoons based on MSLP converted from Vmax using this equation was exactly the same as the number evaluated using the provided MSLP values (Fig. 9a).

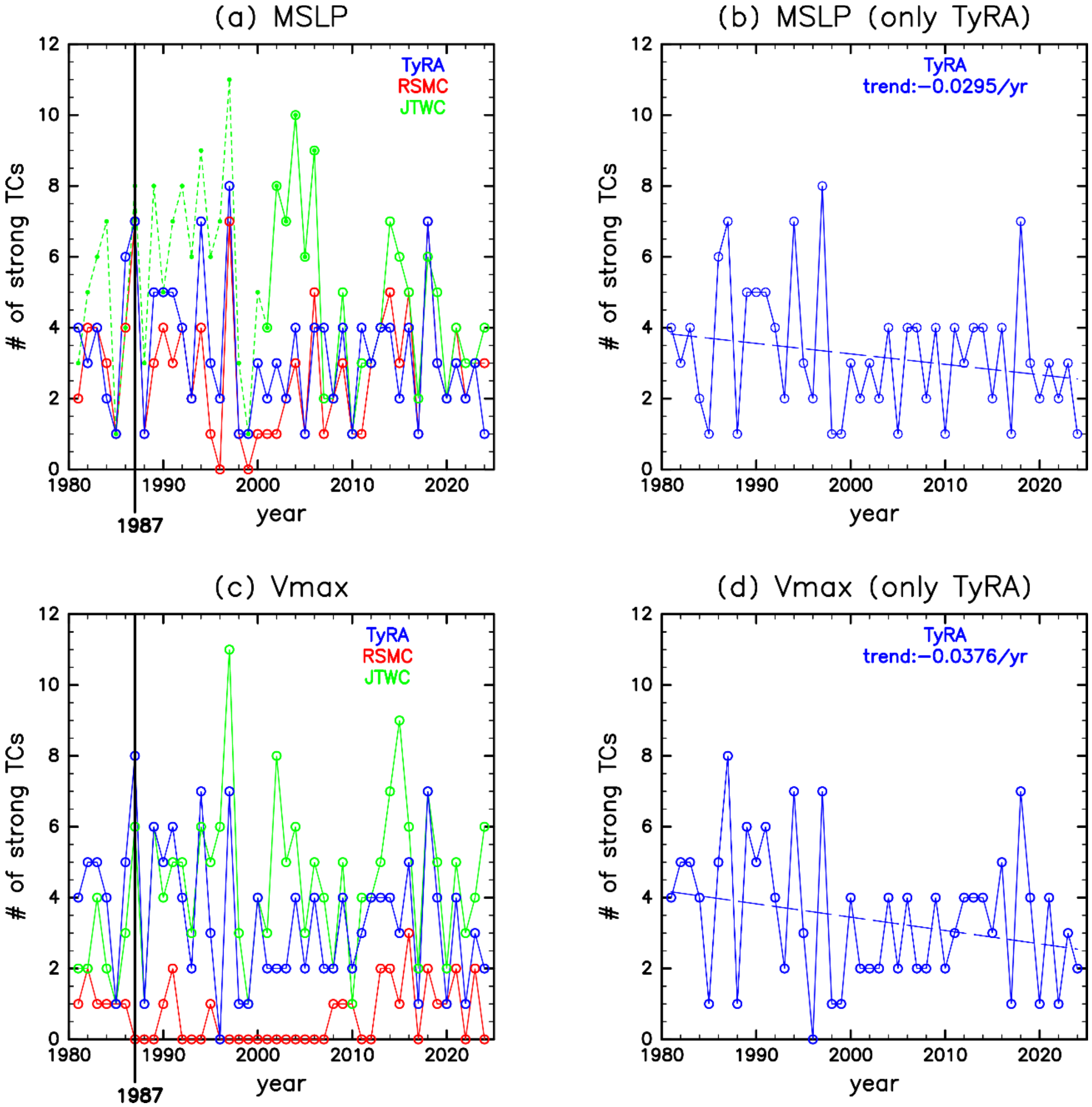

Fig. 9. Number of intense typhoons meeting the thresholds based on (a, b) MSLP and (c, d) Vmax. Details of the thresholds are described in the text. In (a) and (c), TyRA, the RSMC Tokyo best-track, and the JTWC best-track are shown in blue, red, and green, respectively. In (b) and (d), only TyRA is shown. Because complete JTWC MSLP records are available only from 2002 onward, MSLP values converted from Vmax using the relationship of Atkinson and Holliday (1977) are also shown from 1981 to 2006 by the dashed line with small dots.

When the number of typhoons reaching the intense-typhoon criterion based on MSLP is compared, TyRA and the RSMC Tokyo and JTWC best-track datasets show a certain degree of consistency before 1987, when aircraft observations were available, and after the late 2000s (Fig. 9a). In contrast, from 1988, after the termination of aircraft reconnaissance, to the mid-2000s, JTWC records a larger number of intense typhoons, whereas RSMC Tokyo records a smaller number. Recall that Kawabata et al. (2023) reported the discrepancy of the numbers in the intense TCs between the RSMC Tokyo best-track and Dvorak-based estimate during this period. TyRA lies between these two best-track datasets during this period. The long-term trend in TyRA is $-0.0295$ yr$^{-1}$, indicating a slight decreasing trend (Fig. 9b). However, this trend has a p-value of 0.176 and is not statistically significant. This result is consistent with Kawabata et al. (2023). This suggests that, even during the past 40 years under ongoing climate change, the number of intense typhoons has not changed significantly. When intensity is evaluated using 1-min mean Vmax, the number of very intense typhoons in RSMC is small, particularly from the late 1980s to the mid-2000s. From 1987 to around 2000, the numbers of super typhoons in TyRA and JTWC best-track are similar. Thereafter, through around the mid-2010s, the JTWC best-track contains more typhoons reaching the threshold than TyRA. The long-term trend in TyRA is $-0.0376$ yr$^{-1}$, which also indicates a statistically insignificant decreasing trend ($p=0.102$). The total number of typhoon formations over the WNP has itself been decreasing (Hu et al. 2018). When this is taken into account and the values are expressed as the occurrence rate of intense typhoons, the trends are $-0.06\%$ yr$^{-1}$ based on the MSLP criterion and $-0.09\%$ yr$^{-1}$ based on the Vmax criterion. These changes are not statistically significant, with p-values of 0.428 and 0.268, respectively.

Figure 10a shows histograms of typhoon intensity for six-hourly records in TyRA, the RSMC Tokyo best-track, and the JTWC best-track. Compared with the two best-track datasets, TyRA contains slightly fewer typhoons with MSLP around 1000 hPa and slightly more typhoons with MSLP around 950–970 hPa. For intense typhoons with MSLP of 930 hPa or lower, JTWC records more cases than TyRA, whereas RSMC records slightly fewer. Fig. 9 showed that the number of intense typhoons based on MSLP differs substantially among the three datasets from the late 1980s to the mid-2000s, and thus Fig. 10b and 10c show the histograms separately for the first half of the period, 1987–2005, and the second half, 2006–2024. These figures indicate that the larger number of cases with MSLP of 920 hPa or less in JTWC and the slightly smaller number in RSMC occur mainly during the first half of the period. During the second half, the number of records for very intense typhoons does not differ greatly among the three datasets. In addition, the shape of the TyRA histogram changes little between the two periods, suggesting that differences in typhoon intensity analysis methods have had a large influence on the apparent long-term frequency distribution of MSLP over the WNP. Figure 10d–f show the corresponding histograms for 1-min mean Vmax for the full period, the first half, and the second half. For very intense typhoons with Vmax exceeding 120 kt, JTWC records many cases, RSMC records few, and TyRA lies between them, which is similar to the results based on MSLP. In the RSMC Tokyo best-track, very few cases reach 130 kt or higher after conversion to 1-min mean Vmax. Unlike the comparison based on MSLP, differences between the RSMC Tokyo best-track and the other two Vmax datasets remain even during the second half of the period.

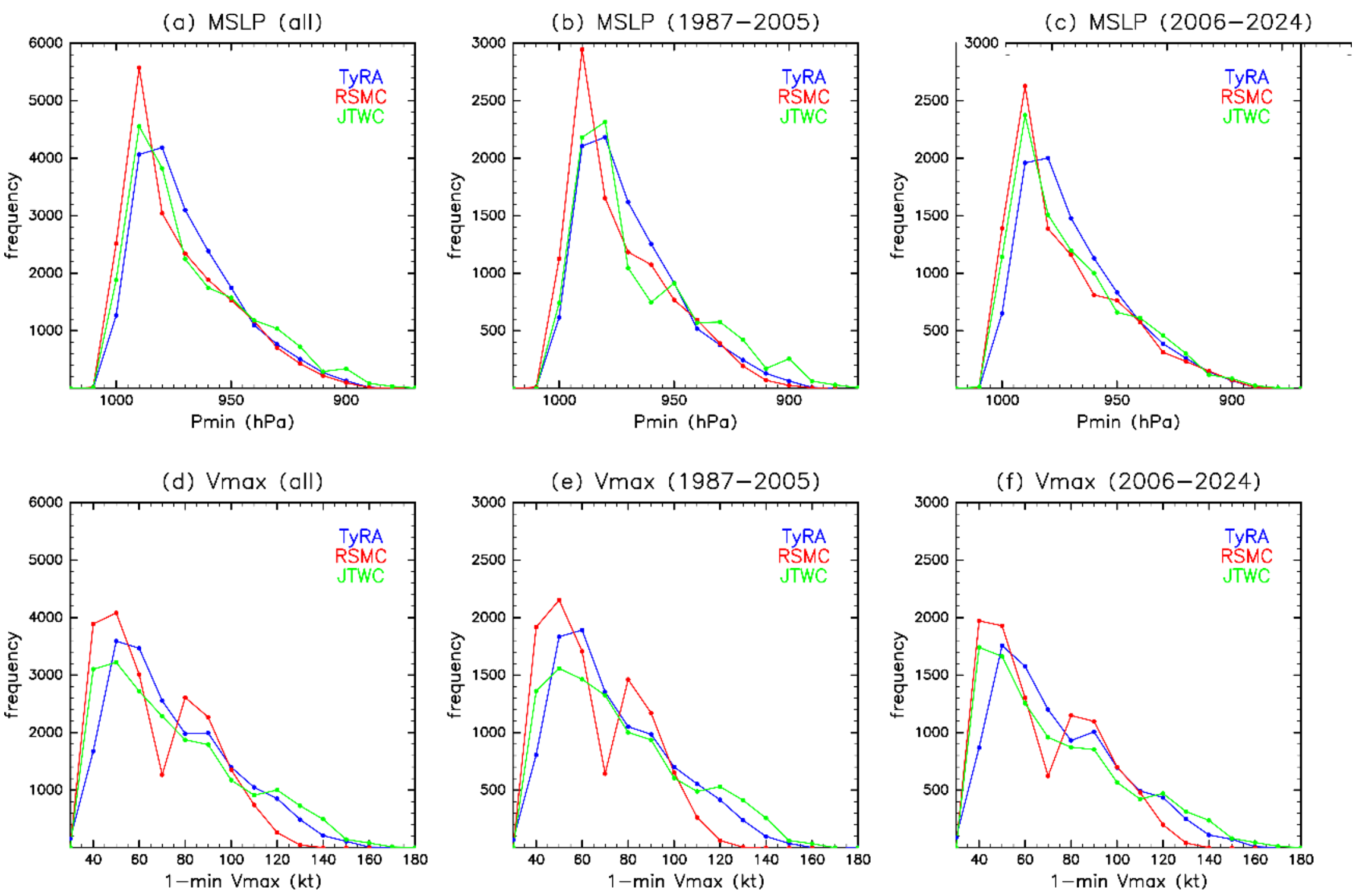

Fig. 10. Histograms of typhoon intensity taking a six-hourly record as one sample: (a–c) MSLP in hPa and (d–f) 1-min mean Vmax in kt. Blue, red, and green indicate TyRA, the RSMC Tokyo best-track, and the JTWC best-track, respectively. Panels (a) and (d) show the full TyRA analysis period from 1987 to 2024, panels (b) and (e) show 1987–2005, and panels (c) and (f) show 2006–2024. For Vmax, the 10-min mean Vmax in the RSMC Tokyo best-track was converted to 1-min mean Vmax.

Figure 11 shows annual mean values of six-hourly MSLP and annual mean values of lifetime minimum MSLP for individual typhoons in TyRA and the RSMC Tokyo and JTWC best-tracks. The figure shows that annual mean MSLP in TyRA tends to be higher than that in JTWC and lower than that in RSMC from 1988 to the mid-2000s. After the late 2000s, the behavior of the annual means is similar among all datasets. The variability component resembles that of the RSMC Tokyo best-track, likely reflecting the fact that both datasets are based on CI numbers analyzed by JMA. Both the annual mean of six-hourly MSLP and the annual mean of lifetime minimum MSLP show slight weakening trends, but these trends are not statistically significant.

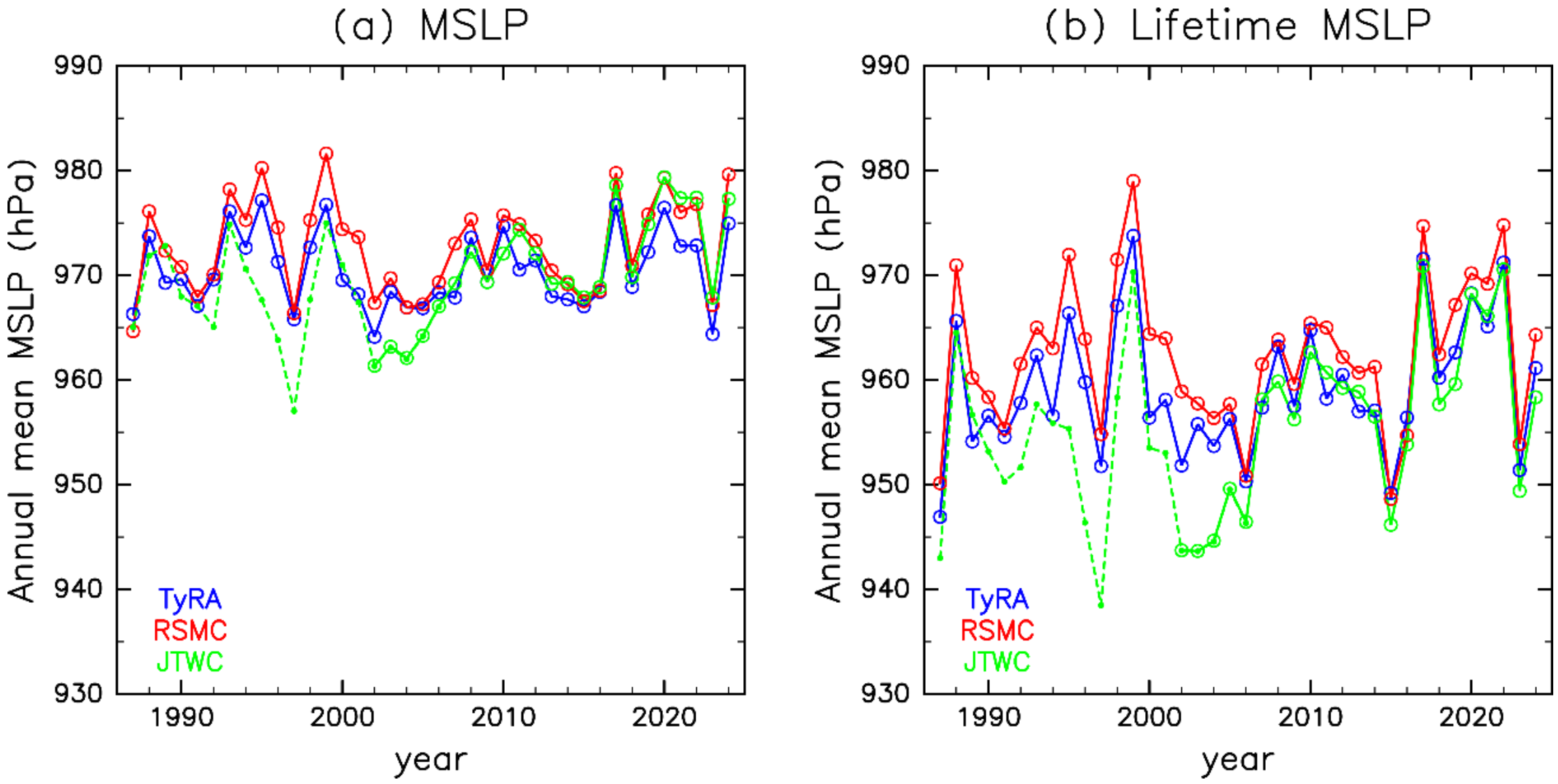

Fig. 11. Annual mean values after 1987 of (a) six-hourly MSLP records and (b) lifetime minimum MSLP for individual typhoons

*3.3 Geographical distribution*

The results described so far indicate that, in TyRA, typhoon intensity over the WNP has shown a statistically insignificant weakening tendency over the past 40 years. Because this is an average feature over the WNP, it is also important to examine regional distributions in intensity and their changes. Figure 12 shows the geographical distribution of mean typhoon intensity. Figures 12a and 12c show low MSLP and large Vmax in the regions around the Philippines, Taiwan, and Okinawa, representing the typical climatological distribution of typhoon intensity over the WNP. Typhoons recorded at low latitudes east of 170°E include cases that crossed over from the central Pacific and had already reached their developing or mature stages. Figures 12b and d show the differences between the first and second halves of the 38-year reanalysis period. East of 130°E, the mean MSLP recently increased while Vmax decreased, indicating a weakening of mean typhoon intensity. This change is particularly pronounced in the southeastern quadrant of the WNP. In contrast, mean typhoon intensity slightly increased west of 130°E, including the South China Sea, the region from east of the Philippines to the southern East China Sea, and the vicinity of Taiwan and Okinawa. The weakening trend in the eastern portion is also evident in terms of CI numbers from JMA (Fig. 12), indicating this trend for the other Dvorak-based intensity estimates.

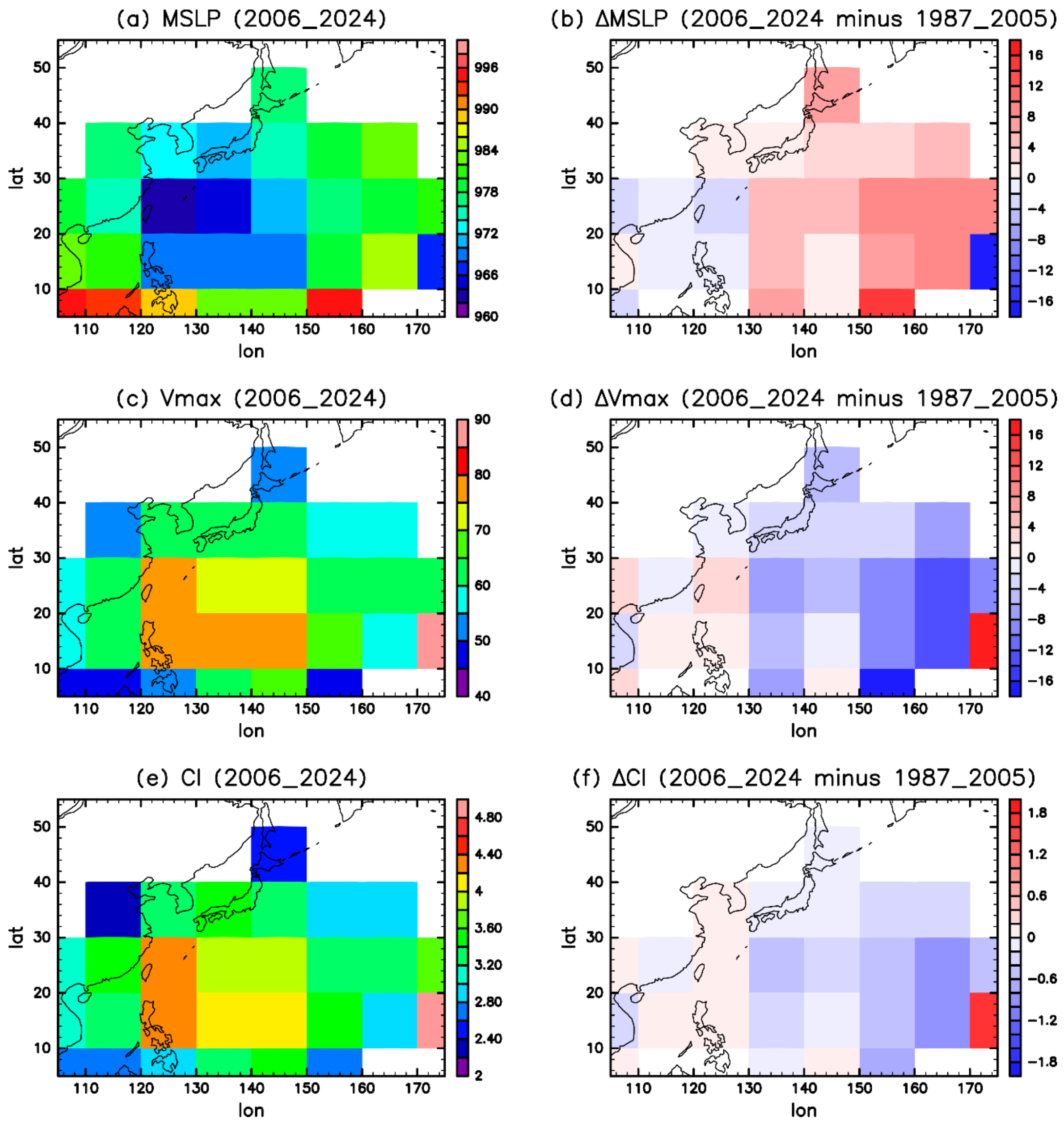

Fig. 12. Geographical distributions of (a) mean MSLP based on six-hourly records during the second half of the reanalysis period, 2006–2024, and (b) the difference in mean MSLP between the second half, 2006–2024, and the first half, 1987–2005. (c, d) Same as (a) and (b) but for the mean Vmax. (e, f) Same as (a) and (b) but for the mean of CI numbers. Mean values are calculated for each 10° × 10° grid box. Grid boxes with fewer than 20 samples are excluded from the plot.

Figure 13 shows mean MSLP and Vmax for each longitude for the early and late halves of the reanalysis period. The results indicate that, in the recent period, mean MSLP increased by approximately 7–8 hPa in the 150°E–170°E longitude range. A *t*-test was applied to the differences in the mean values. Because six-hourly samples are not necessarily independent, the lag-1 autocorrelation coefficient was calculated, and the sample size was replaced by the effective sample size. The results show statistically significant weakening at the 95% confidence level for MSLP in the 150°E–160°E, and for Vmax in the 130°E–140°E and 150°E–160°E.

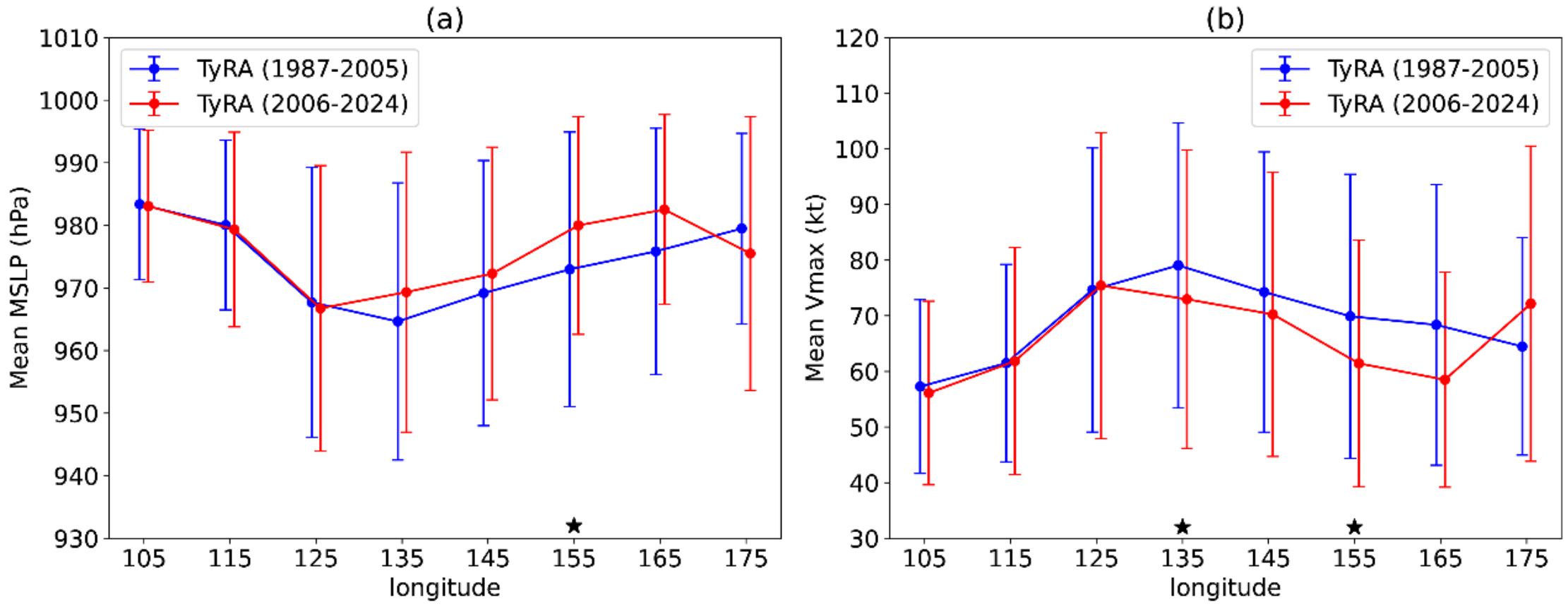

Fig. 13. Mean typhoon intensity of six-hourly records averaged by longitude bands: (a) MSLP and (b) Vmax. Asterisks indicate statistically significant differences at the 95% confidence level. The bars denote the standard deviations.

This weakening trend in the eastern part of the WNP is quite intriguing. One of the potential reasons might be that the locations of typhoon genesis and lifetime maximum intensity have shifted northwestward (Lin and Chan 2015). Because typhoons generally intensify gradually over several days after genesis, the decrease of elapsed time since typhoon genesis can reduce the mean intensity there. Figure 14 shows the longitude and latitude of typhoon genesis locations and locations of lifetime minimum MSLP. The figure indicates that both the genesis locations and the locations of lifetime minimum MSLP have shifted northwestward. To make sure, the mean of typhoon age defined the elapsed time for a record since its genesis was also calculated (Fig. 15). On average, mean age of typhoon records from 130°E to 170°E during 2006–2024 are younger than that during 1987–2005 by 15 h. We also calculated the age by defining the time of TC genesis as the first occasion on which the CI number reached 2.0 or higher, instead of using the RSMC Tokyo best-track record. However, the overall results were not sensitive to the definition of TC genesis (figures not shown).

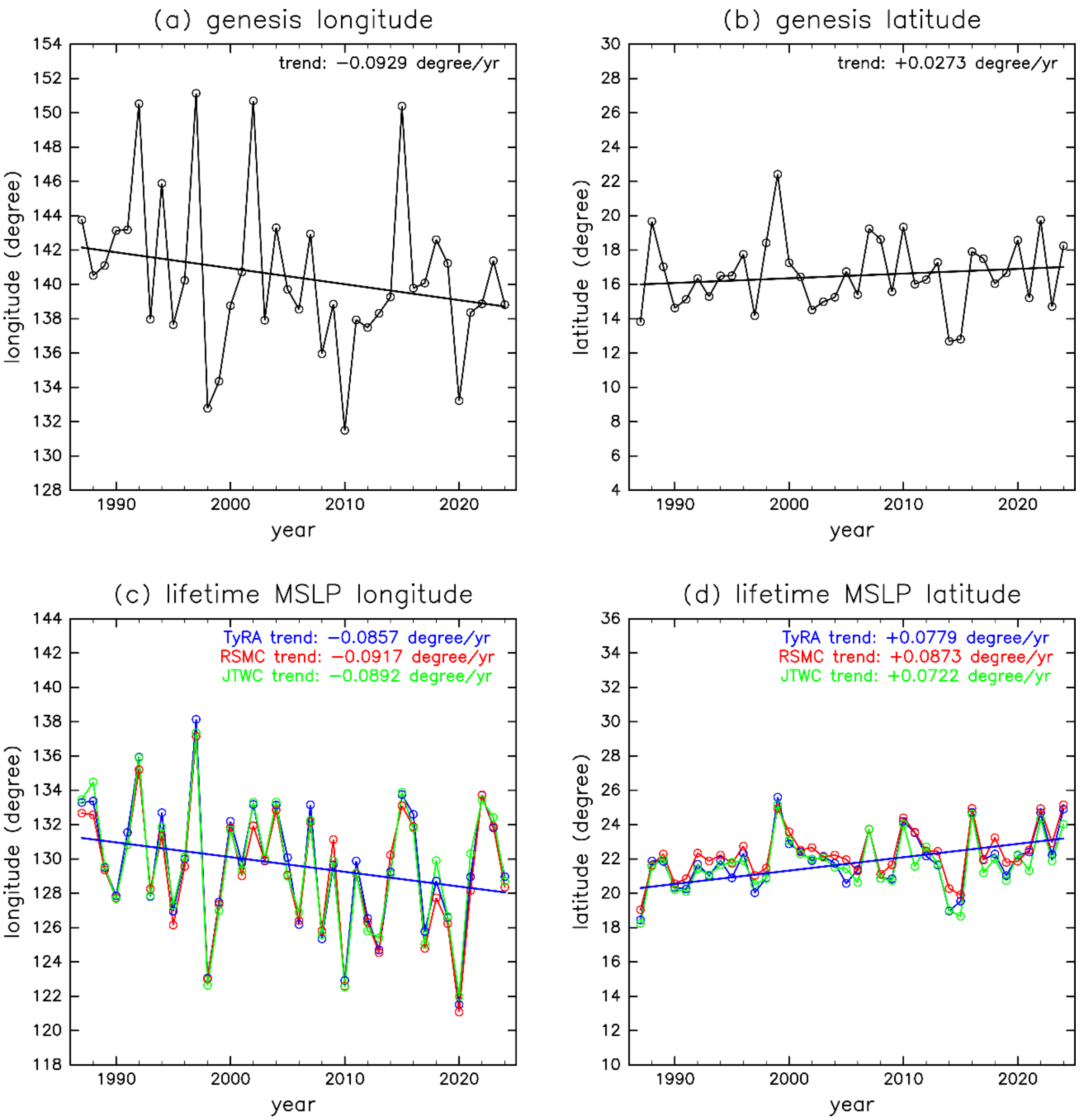

Fig. 14. Annual mean locations of typhoon centers over the WNP: (a) longitude and (b) latitude, and annual mean locations at the time of lifetime minimum MSLP: (c) longitude and (d) latitude. In (a) and (b), the common data used for TyRA and RSMC Tokyo are shown in black. In (c) and (d), results from TyRA, RSMC Tokyo, and JTWC are shown. To avoid visual clutter, regression lines in (c) and (d) are shown only for TyRA.

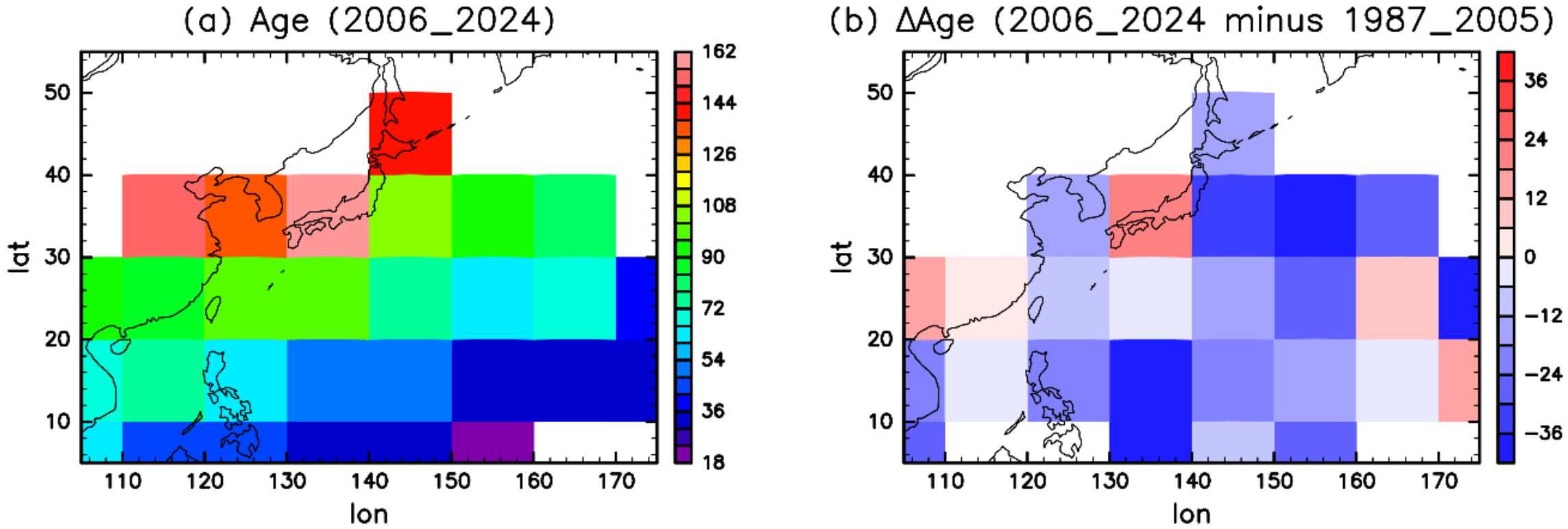

Fig. 15. Geographical distributions of (a) mean age (in h) based on six-hourly records during the second half of the reanalysis period, 2006–2024, (b) the difference of mean ages between the second half, 2006–2024, and the first half, 1987–2005. Mean values are calculated for each 10° × 10° grid box. Grid boxes with fewer than 20 samples are excluded from the plot.

That said, the mean typhoon intensity may weaken in the eastern portion of WNP even under global warming in the past four decades because the typhoons tend to be observed in their early stages of development there. Of course, further research, including investigation of the shift of typhoon geneses and atmospheric and oceanic environmental conditions, is needed to explore the reasons for the change. We are not sure whether this is a signal due to global warming, or whether the impact of global warming is concealed by other long-term natural variabilities. The location shifts may be relevant to a La Niña–like pattern of SST change associated with global warming and/or decadal natural variability such as Pacific Decadal Oscillation and Pacific Meridional Mode (Wang et al. 2015; Watanabe et al. 2024). A detailed examination of the physical mechanisms responsible for the shifts in typhoon locations is left for future work.

*3.4 Rapid intensification*

Figure 16a shows the occurrence rate of RI based on MSLP, treating each six-hourly record as one sample. Here, RI is defined as a decrease in MSLP of 42 hPa or more over the preceding 24 h. Cases within 24 h of genesis are excluded from the count. From the late 1980s to the early 2000s, the RI occurrence rate is largest in JTWC, followed by TyRA, and is much smaller in RSMC. The trend in the number of RI cases in RSMC after 1987 is consistent with Ito (2016) and Shimada et al. (2020). From the late 2000s to the mid-2010s, the three datasets show a certain degree of consistency, although the rapid intensification rate in TyRA is slightly higher. Since the late 2010s, the occurrence rate of RI events has been large in JTWC, followed by RSMC, whereas TyRA records fewer rapid intensification cases, resulting in divergence among the three datasets. TyRA does not show a clear long-term trend. Another notable feature is that the mean age is younger in the region of 10°N–20°N and 130°E–140°E, which is one of the regions with many typhoons intensifying rapidly (Fudeyasu et al. 2018).

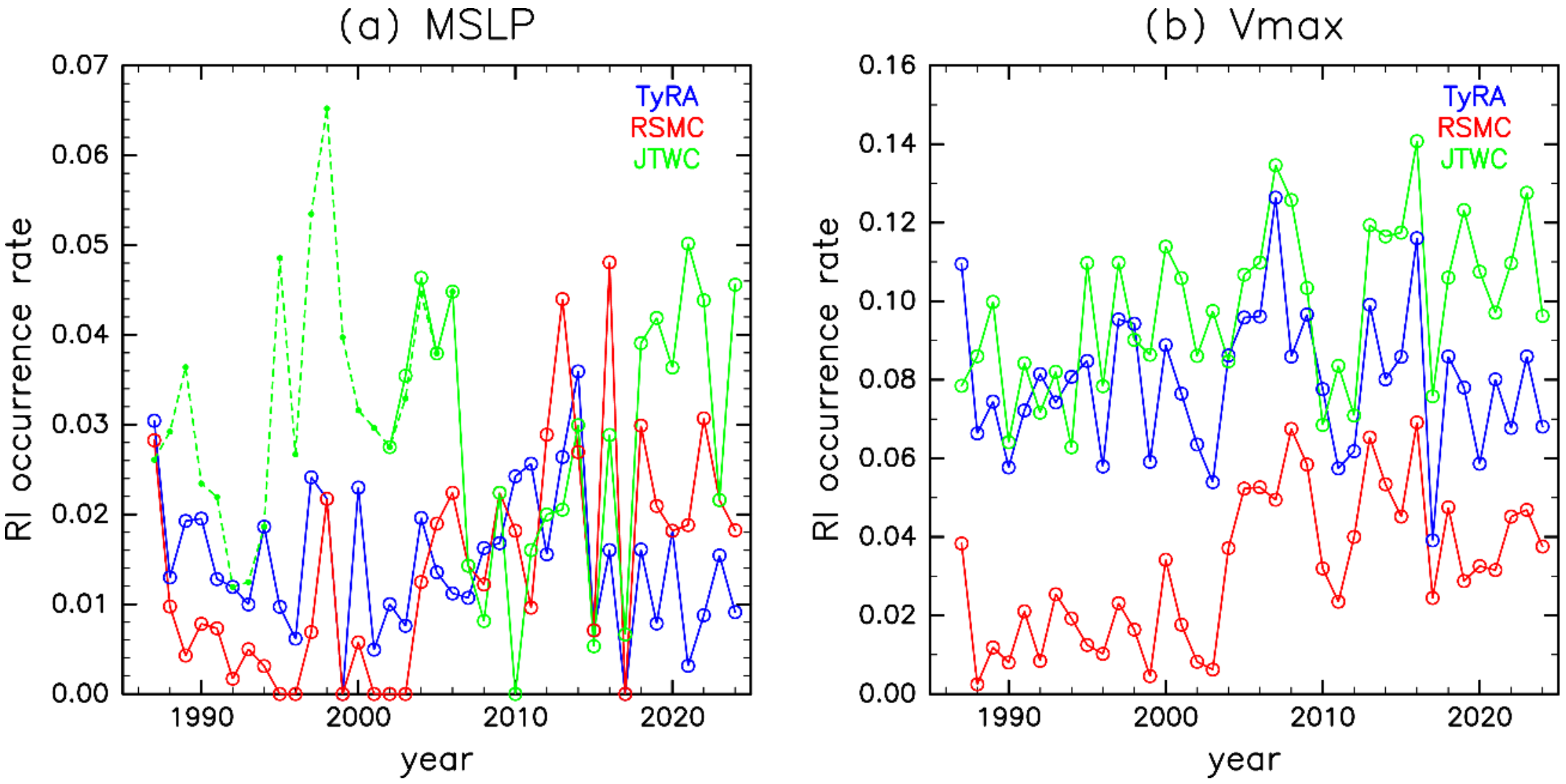

Fig. 16. Occurrence rate of rapid intensification when each six-hourly record is treated as one sample: fractions of cases in which (a) MSLP decreased by 42 hPa or more within 24 h and (b) 10-min mean Vmax increased by 30 kt or more within 24 h.

Figure 16b shows the RI occurrence rate of cases in which Vmax increased by 30 kt or more within 24 h. In this comparison, JTWC and TyRA have higher rates than the RSMC Tokyo best-track. TyRA shows little long-term change, whereas the JTWC and RSMC Tokyo best-tracks show slight increasing tendencies. As described in Section 2.3, the CI numbers used in TyRA were produced with constraints that prevent excessively abrupt changes in T number. Therefore, the actual occurrence rate of RI may be higher than that estimated in TyRA.

*3.5 Evaluation of intensity forecast errors*

Typhoon intensity forecasts are essential for disaster prevention and mitigation. However, typhoon intensity forecast errors are usually measured by treating best-track data as the truth. Therefore, if the quality of best-track data has changed over time, the estimated "intensity forecast errors" may also have been affected by such changes. Ito (2016) reported that the errors in typhoon intensity forecasts by RSMC Tokyo increased around the mid-2000s, and thus it is possible that this increase was associated with changes in RSMC Tokyo best-track quality. We therefore evaluated the interannual variation in RSMC Tokyo intensity forecast errors using both the RSMC Tokyo best-track and TyRA as reference datasets. The evaluation period was 1992–2024, and the typhoon forecast error database of Ito (2016) was used.

Figure 17 shows MSLP and Vmax forecast errors for lead times of 24–72 h. Because the available period for 96–120-h intensity forecasts is short (since 2019), these forecasts are not discussed here. For Vmax, TyRA was converted to 10-min mean values using the conversion described in Eq. (8) and used as the reference. When the RSMC Tokyo best-track is used as the reference, intensity forecast errors gradually increase from the 1990s to the mid-2010s, as reported by Ito (2016). In contrast, when TyRA is used as the reference, the increase in errors from the 1990s to the mid-2010s becomes less clear for 24-h MSLP forecasts and for 24–72-h Vmax forecasts.

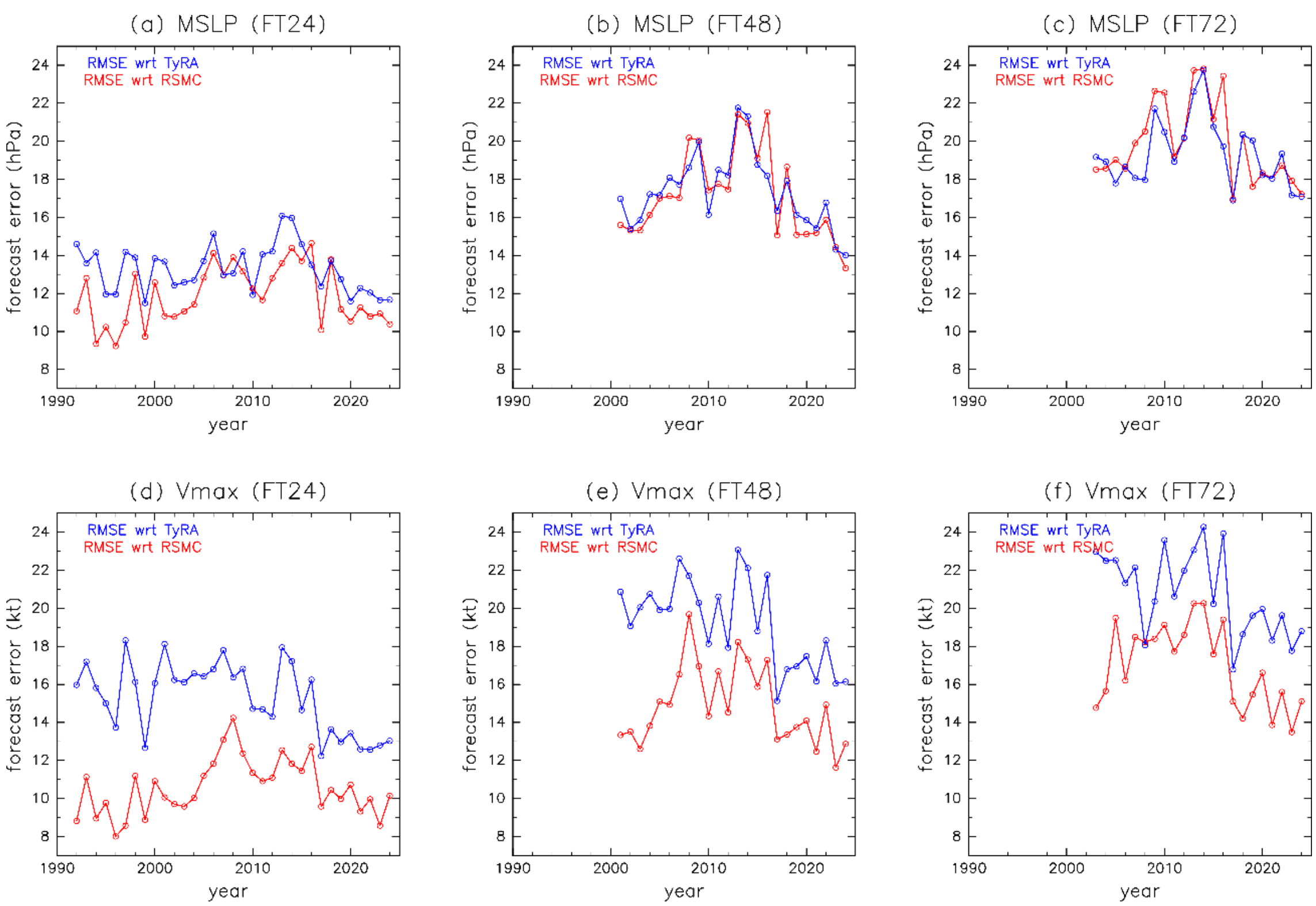

Fig. 17. Errors in typhoon intensity forecasts issued by RSMC Tokyo: (a–c) MSLP and (d–f) Vmax. Panels (a) and (d), (b) and (e), and (c) and (f) show 24-, 48-, and 72-h forecasts, respectively. Because Vmax in TyRA corresponds to a 1-min mean value, it was converted to a 10-min mean value using Eq. (8).

One possible reason for this behavior is a substantial change in the characteristics of rapid intensification events shown in the previous subsection, which tend to be associated with large forecast errors (Ito 2016). During 1992–2017, the correlation coefficient between the 24-h MSLP forecast error and the occurrence rate of rapid intensification was 0.54 in TyRA and 0.75 in RSMC. Thus, the occurrence rate of rapid intensification is closely related to typhoon intensity forecast errors. In the RSMC Tokyo best-track, rapid intensification events were rarely recorded in the 1990s, whereas they became more frequent after the mid-2000s and peaked from the early to mid-2010s, as shown in Fig. 16. In contrast, although TyRA also shows a peak in the rapid intensification rate from the early to mid-2010s, it does not show a pronounced scarcity of rapid intensification events in the 1990s. Thus, the apparently small number of rapid intensification events in the RSMC Tokyo best-track during the 1990s is one reason why RSMC Tokyo intensity forecast errors appeared to increase over time. The high frequency of RI events from the early to mid-2010s is likely another contributing factor.

Since the late 2010s, forecast errors have become somewhat smaller than in the preceding period for both references. This reduction is likely attributable in part to recent efforts to reduce typhoon intensity forecast errors, including the introduction of the new typhoon intensity forecast system in RSMC Tokyo (Yamaguchi et al. 2018).

## 4. Summary and future perspectives

The impact of climate change on typhoons is an issue of great importance. However, best-track datasets, which are commonly used as sources of typhoon intensity information, contain substantial uncertainties, and their quality may have changed over time. In recent years, efforts have been made to construct homogeneous, satellite-based long-term reanalysis datasets of typhoon intensity. Aizawa et al. (2024) demonstrated that further improvements are possible in terms of consistency with aircraft observations over the WNP in the 1980s.

In this study, we conducted a long-term reanalysis of MSLP by combining the Dvorak-method-based CI number reanalysis datasets produced by JMA (Nishimura et al. 2023; Tokuno et al. 2009), recent JMA post-analysis CI numbers, and the relationship proposed by Aizawa et al. (2024), which estimates MSLP accurately from CI numbers and other variables. Specifically, we optimized the relationship against aircraft observations from 1981 to 1986 and applied the resulting Eq. (3) to reanalyze MSLP for 1987–2024. The estimated MSLP values were then converted into 1-min mean Vmax using the relationship of Knaff and Zehr (2007). The resulting dataset of MSLP and Vmax is referred to as the long-term typhoon intensity reanalysis TyRA. Although the conversion relationship proposed by Knaff and Zehr (2007) was used in this study, it has the drawback of being relatively complex and difficult to interpret physically. Chavas et al. (2025) proposed a simpler relationship with good accuracy, and the use of this relationship for such a conversion may be worth considering in future work.

During the optimization period of 1981–1986, the equation used in TyRA provided the best explanation of aircraft-observed MSLP. The Koba relationship also produced reasonable estimates with weak biases. ADT-HURSAT tended to overestimate the intensity of intense typhoons and underestimate the intensity of weak-to-moderate typhoons in terms of MSLP. For recent aircraft-observed MSLP data obtained during T-PARC, ITOP, and T-PARCII, the TyRA and Koba relationships both had RMSDs of approximately 10 hPa. However, further accumulation of observational data is required to evaluate the estimation skill of TyRA, in particular for the intensities of very intense typhoons with MSLP below 920 hPa.

Comparison between TyRA and best-track datasets showed that, from the late 1980s to the mid-2000s, JTWC best-track tended to record a larger number of intense typhoons, whereas RSMC Tokyo best-track tended to record a slightly smaller number. Over the long term, the number of intense typhoons over the WNP as a whole showed a statistically insignificant decreasing trend. The mean typhoon intensity in TyRA also exhibited a slight weakening tendency over the WNP as a whole, but this trend was not statistically significant. In terms of geographical distribution, mean typhoon intensity tended to weaken east of 140°E, particularly in the southeastern quadrant of the WNP. In some longitude bands in the eastern part of the WNP, the results suggested statistically significant weakening of mean typhoon intensity. In contrast, it tended to increase slightly west of 130°E, including the South China Sea, the region offshore of the Philippines, and the vicinity of Taiwan and Okinawa. The gradual northwestward shifts in typhoon genesis locations and locations of peak intensity over the past 40 years may help explain the weakening in the eastern part of the WNP through the increasing fraction of younger typhoons, although further investigation is needed. In any case, these results do not support the interpretation that typhoon intensity has increased over the past several decades as a consequence of global warming. Rather, they indicate that the mean intensity may become weak in the eastern part of WNP.

For rapid intensification, the RSMC Tokyo best-track showed an increasing tendency around the mid-2000s, whereas the occurrence rate in TyRA was relatively stable. In contrast, JTWC exhibited much larger variability. TyRA did not show any pronounced long-term change in rapid intensification. However, because the CI numbers used in TyRA were subject to constraints on abrupt changes in T number, it should be noted that TyRA may underestimate the occurrence rate of rapid intensification while the quality is homogeneous.

Regarding typhoon intensity forecast errors, previous studies reported that forecast errors evaluated against the RSMC Tokyo best-track increased around the mid-2000s. However, when TyRA was used as the reference, this increase was no longer evident for 24-h MSLP forecast errors or for 24–72-h Vmax forecast errors. One possible reason is that typhoon forecast errors are correlated with RI, and the occurrence rate of rapid intensification in the RSMC Tokyo best-track changed substantially around the mid-2000s. This finding highlights the importance of using homogeneous datasets when evaluating the errors in typhoon intensity forecasts.

TyRA was developed using CI numbers with the aim of constructing a dataset that is as accurate as possible under the assumption of long-term homogeneity. Nevertheless, it should be noted that TyRA does not necessarily provide the best typhoon intensity estimate for every individual case. By using the relationship of Aizawa et al. (2024), TyRA aims to achieve typhoon intensity analyses with an MSLP RMSD of approximately 10 hPa. However, when typhoons make landfall, or when ocean buoys or aircraft observations are available, the typhoon intensity analyses in the RSMC Tokyo best-track may be closer to the true values than the estimates obtained in this study, which are based primarily on CI numbers. In the future, it may be possible to correct TyRA values when in situ observations are available. It should also be recognized that, because TyRA is based on a regression equation, it extracts the mean state corresponding to a given CI number, development rate, latitude, storm size, and environmental pressure. In other words, extreme values arising from variations in actual typhoon intensity that cannot be explained by the regression equation may have been removed. In addition, although the resolution of infrared satellite imagery in the 1980s was approximately 5 km, caution is still required for cases such as typhoons with very small eyes. Also, we note that precipitation associated with typhoon in WNP may have changed in a different way as a result of global warming and other natural variability in the last four decades (Chen et al. 2025).

Despite these limitations, TyRA is a new long-term typhoon intensity reanalysis product that is consistent with aircraft observations obtained over the WNP in the 1980s, and it has considerable potential for a wide range of applications. TyRA is expected to be useful for more detailed studies of climatological changes in typhoons, as bogus data for developing new global reanalysis datasets, as fundamental data for more accurately examining the relationship between atmosphere and ocean environmental conditions and typhoon intensity, as a reference dataset for hindcast evaluation, and as input data for ocean and wave hindcast simulations.

*Acknowledgments*

We thank Dr. John Knaff, Dr. Chun-Chieh Wu, Dr. Robert Rogers, and Dr. Masuo Nakano for their thoughtful discussions. This study was supported by the General Collaborative Research Program of the Disaster Prevention Research Institute, Kyoto University (2025-GC06), MEXT-program SENTAN (JPMXD0722678534), the Moonshot Research and Development Program, Goal 8 (JPMJMS2282-06), and JSPS KAKENHI Grant-in-Aid for Scientific Research (S) (JP21H04992). The opinions in this paper are those of the authors and should not be regarded as official views of JMA.

*Data availability*

The TyRA dataset, the research product developed in this study, is available from the Kyoto University Research Information Repository (http://hdl.handle.net/2433/301586). Aircraft reconnaissance data for typhoons in the 1980s are documented in the JTWC Annual Reports and the Monthly Weather Reviews of the Japan Meteorological Agency. The T-PARC and ITOP data were obtained from the NSF NCAR Earth Observing Laboratory (EOL) data archive (https://data.eol.ucar.edu/). The T-PARCII data were distributed through the Global Telecommunication System (GTS), and more detailed data are available upon request from tsuboki@nagoya-u.jp. The reanalyzed and operational CI number were provided by the RSMC Tokyo. The RSMC Tokyo and JTWC best-track datasets are available at https://www.jma.go.jp/jma/jma-eng/jma-center/rsmc-hp-pub-eg/trackarchives.html and https://www.metoc.navy.mil/jtwc/jtwc.html?best-tracks, respectively. The JRA-3Q dataset is available at https://www.data.jma.go.jp/jra/html/JRA-3Q/index_en.html. The typhoon intensity forecast error database is available at https://ssrs.dpri.kyoto-u.ac.jp/itokosk/distribution_e.html.